\documentclass[12pt]{article}
\usepackage[cp1251]{inputenc}
\usepackage[english]{babel}
\usepackage{titlesec}
\usepackage[titletoc,title]{appendix}

\usepackage{amsfonts,epsfig,epstopdf}
\usepackage{amssymb,amsmath,enumerate,float}
\usepackage[section]{placeins}
\usepackage{xcolor}
\newcommand{\redtext}{\textcolor{black}}
\makeatother
\addto\captionsenglish{}

\renewcommand{\i}{i}
\newcommand{\ptl}{\partial}

\newcommand{\be}{\begin{equation}}
\newcommand{\ee}{\end{equation}}
\newcommand{\beq}{\begin{equation}}
\newcommand{\eeq}{\end{equation}}

\newcommand{\rmd}{ d}
\author{M.A.~Mironov, A.V.~Shanin, A.I.~Korolkov, K.S.~Kniazeva}
\title{\redtext{Transient processes in a gas / plate structure in the case of light gas loading}}
\begin{document}
\maketitle
\begin{abstract}

Problems of pulse excitation in an acoustic waveguide with a flexible wall 
and in an acoustic half-space with a flexible wall are studied. In both cases the 
flexible wall is described by a thin plate equation. The solutions are 
written as double Fourier integrals. The integral for the waveguide is 
computed explicitly, and the integral for the half-space is estimated asymptotically.  
\redtext{A special attention is paid to the pulse, which is 
a harmonic wave of a finite duration associated with the coincidence point of the 
dispersion diagrams of the acoustic medium and the plate. 
The method of estimating of the double Fourier integral is applied to the problem of 
excitation of waves in a system composed of an ice plate, water substrate, and the air.} 

\end{abstract}

%\begin{keyword}
%phase synchronism\sep dispersion equation\sep analytical continuation\sep elastic plate 
%\end{keyword}

%\linenumbers
%%%%%%%%%%%%%%%%%%%%%%%%%%%%%%%%%%%%%%%%%%%%%%%%%%%%%
\section{Introduction}

The paper is motivated by a simple experiment. 
A thin ($\sim 3$~cm) layer of ice on a lake or a pond should be hit   
with a stick or stone. Let a listener be located on the surface of the lake quite far from the 
place where the ice is hit (about 200~m). The listener shall hear a long quasi-monochromatic 
``whistle'' instead of a kick. The sound is surprising and unexpected. 
The current paper describes such a signal mathematically. 

%The signal is referred to as a pulse related to the 
%a {\em coincidence point pulse\/} in this paper (the reason ).  

The qualitative explanation of the effect is rather simple. 
The impact on the ice generates a wide spectrum of bending waves, 
propagating along the ice plate.
The bending oscillation of the ice is then a source of the sound 
that is recorded in the air. 

It is important that the bending waves possess 
{\em dispersion}. 
The higher is the frequency, the greater is the phase velocity. 
At some circular frequency $\omega_*$, that is called the {\em coincidence frequency}, the phase velocity of the wave coincides with the air sound velocity~$c$. 
At this frequency, the acoustic wave propagating in a sliding way along the ice 
has the same phase velocity as the bending wave, 
and the energy transfer from the ice plate to the sound wave is the most effective.
One can show by measurements that the pulse related to the coincidence point 
has a narrow--band 
spectrum centered at the coincidence frequency. 

As usual, the dispersion of the bending waves in the ice leads to 
emerging of the concept of the 
{\em group velocity}, i.~e.\ of the velocity of narrow-band pulse propagation. 
For a bending plate without loading, the relation between the wavenumber $k$
and the temporal frequency $\omega$ is $k \sim \sqrt{\omega}$, thus the 
group velocity is two times bigger than the phase velocity:
\[
\left( \frac{\rmd k }{\rmd \omega} \right)^{-1} = 2 \frac{\omega}{k}.
\]
\redtext{
A detailed study (see below) shows that 
the air and water loading does not change this situation qualitatively. 
The group velocity $v_1$ of the bending waves at the coincidence frequency 
is equal to $2c$ if the water substrate is ignored, and 
can be roughly estimated as
$2.3c$ if the water is taken into account. 
The waves in the air going horizontally are dispersionless, thus
the group velocity $v_2$ for them is equal to the phase velocity~$c$:
$v_2 = c$. }

\begin{figure}[ht]
\centerline{\epsfig{file=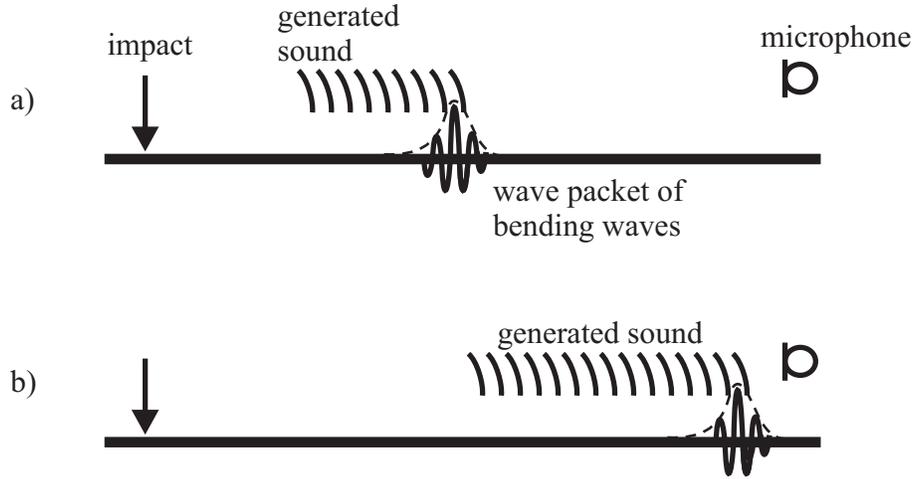, width= 12cm}}
\caption{To the explanation of the ice/air experiment}
\label{fig00}
\end{figure}

Immediately after the impact, the components of the spectrum at the coincidence frequency start to radiate sound into the air. As the group velocity of bending waves 
is about twice as much as the air waves, the wave packet bearing the radiating harmonics propagates to the observation point, permanently leaving behind the sound radiated earlier in the air (see Fig.~\ref{fig00},~a).

The sound signal appears at the observation point only when the wave packet of bending waves comes to this point (see Fig.~\ref{fig00},~b). After that, the sound radiated earlier by the distant points comes. 
The sound signal vanishes only when the sound from the start point comes.
  
The pulse duration is equal to the difference between arrival time of the head of the bending waves packet $L / v_1$ at the coincidence frequency and the arrival time of the sound wave of the impact itself $L/ v_2$:
\[
 L \left( \frac{1}{v_2} - \frac{1}{v_1} \right) .
\]

%%%%%%%%%%%%%%%%%%%%%%%%%%%%

The experimental setup described above is still rather complicated for 
analytical modeling. We simplify it by making the geometry two--dimensional and by eliminating the 
water layer underlying the ice, i.~e.\ only the plate (ice) and the gas (air) 
are considered. 
Moreover, we assume that the air loading is light, i.~e.\ that the 
air wave does not affect the process in the plate.  

Two physical situations are considered in the paper. The first one is  a
thin waveguide with a rigid wall and a flexible wall. 
The second physical situation is a flexible plate loaded by gas half-plane on one side. 
We are mostly interested in the second problem, however the first one can be solved analytically, and thus it can be used to check the statements made for the second one. 

For each formulation, 
the wave excitation problem can be solved formally by the Fourier transform. 
As the result, the wave pulse   
becomes described as a 2D Fourier integral. 
The aim of this paper is to obtain an asymptotic estimation of this 
integral. 

The assumption of light gas loading simplifies the consideration a lot.
The physical system becomes split into two subsystem: the heavy one (the plate), and the 
light one (the gas). 
The denominator of the integrand of the Fourier integral is expressed as a 
product of two functions, each of which can be treated as a dispersion function 
for an isolated subsystem. The zero sets of the dispersion functions are branches of
the {\em dispersion diagram\/} of the system. The branches of the
dispersion diagram are crossing. It is known that if the interaction between the 
subsystems is not negligible, there exists a so-called avoiding crossing of the branches 
instead of a crossing. An estimation of a 2D Fourier integral with singularities 
having an avoiding crossing seems to be a more complicated task. 

\redtext{
Beside two model problems, we formulate a considerably more complicated problem corresponding to the 
experiment with the ice layer. However, we demonstrate that the methods developed in the 
paper can be applied to this problem as well. Moreover, we show that the experimental 
signal displays some properties that follow from our analysis.  
}

%The acoustic field in the air should be found. 
%Only the case of light air loading is considered. The problem is solved rigorously, but then %the limit of small air density is taken. This simplifies the integrals in a high degree. 

The interaction of waves in an elastic plate and a 
surrounding liquid or gas is a well-studied topic. 
This is an interesting analytical problem 
having important practical 
applications. Rather than writing a comprehensive review of the subject, we mention here only the papers important for the research below.  

A 3D problem of a point source time--harmonic excitation of a plate loaded by a liquid 
was formulated and solved by using the Fourier--Bessel transformation  
in~\cite{Tamm1946}. The energy carried by sonic waves was computed. 
Besides, the formal solution of the same problem 
in terms of the Fourier integral can be found in \cite{Morse1968,junger1986sound}. 
A basic analysis of the integral from \cite{Tamm1946} was performed in 
\cite{gutin1965} with the help of the saddle point method. 
Particularly, an asymptotic estimation of the Fourier integral in the far field zone  for the frequencies below the coincidence frequency was found. 

``Free waves'' corresponding to the poles of the Fourier integral are discussed
in \cite{Crighton1979,Rokhlin1989,Freedman1995}. These poles form a complicated structure if the 
loading of the plate cannot be considered as light: they obey an equation of degree~5 even for the simplest plate equation.  

A thick plate  (elastic layer) immersed in the fluid was studied in \cite{Rokhlin1989,Freedman1995,Sorokin2007}. Plates governed by  sophisticated  equations of motion were considered in \cite{Feit1966,Stuart1976,Stuart1976a,Smith2007}. 
A Timoshenko--Mindlin plate loaded by a layered medium was studied in 
\cite{Kurtepov1970}. All such problems can be addressed 
by the techniques described in \cite{Brekhovskih1980}.

In \cite{DiPerna2001, DiPerna2003} the dispersion equation was approximated by a rational function. The acoustical  radiated field was calculated under this approximation. 

A systematic study of heavy/light loading of plates and membranes 
was performed in \cite{Crighton1979,Crighton1980,Crighton1983,
Crighton1984a,
Crighton1992}. 
The importance of the parameter describing the lightness of the loading was stressed 
(this parameter tends to zero in our study below). The regimes were classified. The results were summarised in the 1988 Rayleigh medal lecture by Crighton \cite{Crighton1989}.
An asymptotic expression for a leaky wave in the case of a lightly loaded plate at frequencies above the coincidence was presented in \cite{Crighton1979}.
A lightly loaded membrane excited by a concentrated force was studied
in \cite{Crighton1983}.  
This paper should be mentioned especially because 
the author built a ``map of asymptotics'' covering the whole range of the parameters. 
In \cite{Crighton1984a} some asymptotic results for a heavily loaded plate were obtained.
In \cite{Chapman2005} an intermediate regime of loading was studied, when the fluid loading was heavy enough to affect the surface waves, but the inertia of the plate still 
could be considered as negligible.

Most of the papers (and almost all above) dealt with a time-harmonic wave excitation. 
Only a few considered a transient regime.
The reason for this is the absence of well established methods for
asymptotic estimation of 2D Fourier integrals. 
Note that we develop such methods, but only for the simplest case of  
light loading. 

A formal solution of the transient problem in an integral form 
was obtained in \cite{Stuart1976,Chapman2005}.  
Some works devoted to transient processes in fluid--loaded elastic plates are 
\cite{Magrab1968, Stuart1972}. 
The paper most close to the current study was
\cite{Mackertich1981}, where a series of contour deformations was performed.
The 2D problem was studied there.  
An expression for a first arrival pulse was obtained. In
\cite{James1987,Scherrer2013,Langlet2014} transient processes were studied numerically.   

\redtext{The pulse 
related to the coincidence point of the dispersion diagram was considered 
in \cite{Akolzin2001}. The  energy of the pulse was estimated.} 

The current paper is organized as follows. In Section~2, the problems of wave excitation in a 
waveguide with a flexible wall and in a half-space with a flexible wall 
are formulated and solved by the Fourier transformation. 
\redtext{
Also we formulate a 3D problem for a realistic 3D configuration of an ice layer loaded by air a
and by a water substrate. }
In
Section~3, a solution for the waveguide
with an flexible wall is computed rigorously by a residue integration.   
In Section~4, the integral for the half-space with a flexible wall is estimated. 
According to the ``main statement'' of asymptotic estimation formulated in this section,  
only the crossings of branches of the dispersion diagram and the saddle points on the dispersion diagrams should be taken into account.
As a result,  
a ``library'' of asymptotics related to different fragments of the dispersion diagram is developed. 
\redtext{
In Section~5, we briefly comment on the 3D problem with water loading. Namely, we 
the influence of the water substrate, show how our methods can be applied to the 
Fourier--Bessel integral, comment on the system with absorption, 
and propose an interpretation of the experimental results.  
}
In Appendix, the standard integrals used for the asymptotic study are described. 

%%%%%%%%%%%%%%%%%%%%%%%%%%%%%%%%%%%%%%%%%%%%%%%%%%%%%
\section{Formulation of the problems and integral representations of the field}

%%%%%%%%%%
\subsection{Problem 1. Gas / plate waveguide}

\begin{figure}[ht]
\centerline{\epsfig{file=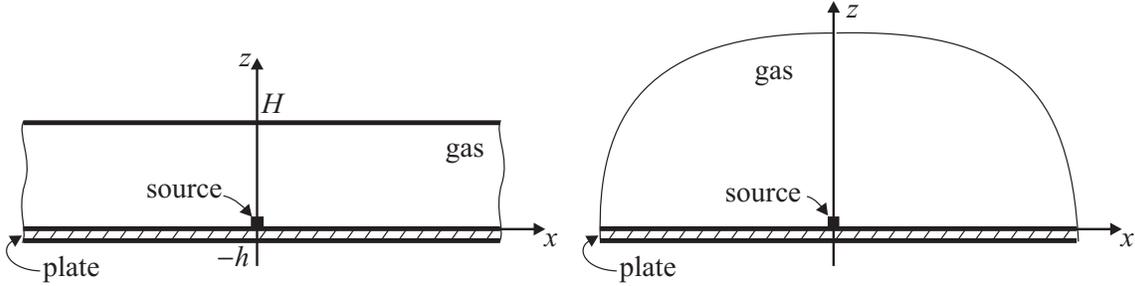, width = 15cm}}
\caption{Geometry of Problem~1 (left) and Problem~2 (right)}
\label{fig01}
\end{figure}

The geometry of the problem is as follows. 
A gas layer occupies the domain  
$0 < z < H$ in the $(x,z)$-plane (see Fig.~\ref{fig01}, left). An elastic plate made of an isotropic material is the layer 
$-h < z < 0$. Assume that $h$ is small and that the bending waves in the plate  
are described by the linear thin plate theory \cite{Landau2004}. 

The plate is stress-free at the surface $z = -h$, the contact conditions are fulfilled 
at the surface $z = 0$, and the surface $z = H$ is acoustically hard (impenetrable).
The source is applied to  the plate at the point $(0, 0)$. 
The time profile of the source is the Dirac's delta-function.

%%%%%%%%%%%%%%%%%%%%%%%%%%%%
%%%%%%%%%%%%%%%%%%%%%%%%%%%%%%%%%%%%%%% 

Describe waves in the gas by an acoustic potential $\phi (t, x, z)$.
The acoustic potential is linked with the pressure $p$ and the particle velocity 
components $(v_x , v_z)$ by the relations \cite{Morse1968}
\begin{equation}
p = \rho \frac{\ptl \phi}{\ptl t},
\qquad 
v_x = - \frac{\ptl \phi}{\ptl x},
\qquad 
v_z = - \frac{\ptl \phi}{\ptl z},
\label{eq1001}
\end{equation}
where $\rho$ is the (constant) gas density. 
The governing equation in the gas is the wave equation 
\begin{equation}
\left( 
\frac{\ptl^2 }{\ptl x^2 } + 
\frac{\ptl^2 }{\ptl z^2 } - 
\frac{1}{c^2} \frac{\ptl^2 }{\ptl t^2 }
\right) \phi(t,x,z) = 0 ,
\label{eq1002}
\end{equation}
where $c$ is the (constant) wave velocity in the gas. 

Let $\zeta (t , x)$ be the vertical displacement of the plate. 
The equation of motion of the plate can be written as \cite{Landau2004}
\begin{equation}
\left( 
T \frac{\ptl^4}{\ptl x^4} 
+ 
\rho_p h \frac{\ptl^2}{\ptl t^2} 
\right) \zeta(t,x) + p(t,x,0) = - f_0\delta(t) \delta(x).
\label{eq1003}
\end{equation}
Here $\rho_p$ is the density of the plate, 
\begin{equation}
T = \frac{E h^3}{12 (1-\nu^2)}
\label{eq1004}
\end{equation}
is the flexural stiffness of the plate ($E$ is the Young's modulus of the
elastic material, and $\nu$ is the Poisson's ratio), $f_0$ is the amplitude 
of excitation.

The second term in the left-hand side of Eq.~(\ref{eq1003}) is responsible for the 
gas loading, and the right-hand side corresponds to the point source excitation. 

The boundary conditions for the air layer are 
\begin{equation}
\left. \frac{\ptl \phi }{\ptl z} \right|_{z = H}= 0,
\label{eq1005a}
\end{equation}
\begin{equation}
\left. 
\frac{\ptl \phi}{\ptl z}
\right|_{z = 0}
 = - \frac{\ptl \zeta}{\ptl t}.
\label{eq1005}
\end{equation}
 
Our aim is to find the pressure in the gas near the plate, i.~e.\ the function  $p(t, x, 0)$.
The solution should be causal, i.~e.\ the field components 
$\phi$ and $\zeta$ should be equal to zero for $t < 0$.

To solve the problem formulated as Eqs~(\ref{eq1002}-\ref{eq1005}), introduce the 
Fourier transform with respect to $x$ and the Laplace transform with respect to $t$:
\begin{equation}
\tilde w(\omega , k)
= \frac{1}{4\pi^2} 
\int \limits_{-\infty}^{\infty}
\int \limits_0^{\infty}
w (t, x) \exp\{ -\i k x + \i \omega t\} \, \rmd t \, \rmd x,
\qquad
{\rm Im}[k]  =0, \quad {\rm Im}[\omega] \geq 0 .
\label{eq1006}
\end{equation} 
Note that the variable of the Laplace transform is 
chosen to be $- i \omega$, so the resulting notations are ``Fourier-like''. 
The inverse of this transform is
\begin{equation}
w(t , x)
= 
\int \limits_{-\infty}^{\infty}
\int \limits_{-\infty + \i \epsilon}^{\infty + \i \epsilon}
\tilde w (\omega, k) \exp\{ \i k x - \i \omega t\} \, \rmd \omega \, \rmd k,
\label{eq1007}
\end{equation} 
where $\epsilon$ is an arbitrary positive parameter. 

Using these transformations, one can write down the acoustic pressure 
near the plate
in the form of a 
double integral  
\begin{equation}
p(t, x, 0) = -\frac{\rho f_0}{4 \pi^2} 
\int \limits_{- \infty}^{\infty}
\int \limits_{- \infty + \i \epsilon}^{\infty + \i \epsilon}
\frac{ \omega^2 \cos (\gamma H)
\exp\{ \i k x - \i \omega t \}
}{\gamma \sin(\gamma H) \, 
(T k^4 - \rho_p h \omega^2) + \omega^2 \rho \cos(\gamma H)}
\rmd \omega \, \rmd k,
\label{eq1008}
\end{equation}
where 
\begin{equation}
\gamma = \gamma (\omega , k) = \sqrt{\omega^2 / c^2 - k^2}.
\label{eq1009}
\end{equation}
The value of the square root Eq.~(\ref{eq1009}) is selected in such a way that 
it has a positive imaginary part on the whole integration plane.
This choice provides existence of only decaying waves in the air. Also, it leads to continuous $\gamma$ on the integration plane.

%The value $\epsilon$ can be chosen arbitrarily. However, if Eq.~(\ref{eq1008})
%is used for practical calculations, one should take it in such a way 
%that the value $\epsilon t$ is not large. If $t$ is estimated as $\sim x / c$, 
%then the value $\epsilon x / c$ is not large. Thus, for further 
%estimations we assume that 
%\begin{equation}
%\epsilon \sim c/ x. 
%\label{eq1009a}
%\end{equation}
%This value will be used in the local evaluation of the integral.  

Let us make two simplifications of Eq.~(\ref{eq1008}). 
First, let $\rho$ be small, i.~e.\ let the gas loading of the plate be light 
(some discussion  
can be found in \cite{Crighton1979,Crighton1984a,Chapman2005}). 
The second term in the denominator can be neglected comparatively to the first one on the 
whole integration plane: 
\begin{equation}
p(t, x, 0) \approx -\frac{\rho f_0}{4 \pi^2 T} 
\int \limits_{- \infty}^{\infty}
\int \limits_{- \infty + \i \epsilon}^{\infty + \i \epsilon}
\frac{\omega^2 \cos (\gamma H)
\exp\{ \i k x - \i \omega t \}
}{(k^4 - \sigma \omega^2) \gamma \sin(\gamma H) }
\rmd \omega \, \rmd k,
\label{eq1010}
\end{equation}
where
\begin{equation}
\sigma = \rho_p h / T.
\label{eq5001a}
\end{equation}

The second simplification for Eq.~(\ref{eq1008}) 
is to assume that the air waveguide is narrow, i.~e.\ 
to take $|\gamma H| \ll 1$. 
%Indeed, this assumption can be valid only 
%in some finite domain of the $(\omega, k)$-plane. So, the output signal 
%is assumed to be filtered by a low-pass filter with a sufficiently high cut-off frequency. 
The assumption yields
$\cos (\gamma H) \approx 1$, $\sin (\gamma H) \approx \gamma H$, and
\begin{equation}
p(t, x, 0) \approx  \frac{\rho f_0}{4 \pi^2 T H} 
\int \limits_{- \infty}^{\infty}
\int \limits_{- \infty + \i \epsilon}^{\infty + \i \epsilon}
\frac{\omega^2 
\exp\{ \i k x - \i \omega t \}
}{(k^4 - \sigma \omega^2)(k^2 - \omega^2 / c^2) 
 }
\rmd \omega \, \rmd k .
\label{eq1012}
\end{equation}
Physically, 
this assumption means that 
only the piston mode is allowed to propagate in the acoustic part of the waveguide.

\redtext{
The resulting representation Eq.~(\ref{eq1012}) 
is one of the simplest integrals possessing a 
pulse related to the coincidence point in its asymptotics. 
}

%Thus, we represented the field in the form of the integral 
%\begin{equation}
%p(t, x, 0) = \frac{\rho }{4\pi^2 H}
%\int \limits_{- \infty}^{\infty}
%\int \limits_{- \infty + i \epsilon}^{\infty + i \epsilon}
%\frac{ \omega^2 
%\exp\{ i k x - i \omega t \}
%}{ D_1 (\omega, k) 
%D_2 (\omega , k) }
%d \omega \, d k,
%\label{eq1012a}
%\end{equation}
%with 
% the new dispersion function for the air layer:
%\begin{equation}
%D_2(\omega) = \omega^2 / c^2 - k^2. 
%\label{eq1012b}
%\end{equation}
%This new function corresponds to a new dispersion diagram 

The denominator of Eq.~(\ref{eq1012}) has two factors that are related to the plate 
and to the air piston mode. 
Introduce the {\em real dispersion diagrams\/}
\begin{equation}
d_1 = \{ (\omega , k) \in \mathbb{R}^2  \, | \, k^4 - \sigma \omega^2 = 0 \},
\qquad 
d_2 = \{ (\omega , k) \in \mathbb{R}^2 \,  | \, k^2 - \omega^2 / c^2 = 0 \}. 
\label{eq1011b}
\end{equation}
Introduce also the 
{\em complex dispersion diagrams\/} $d_1^{\rm c}$, $d_2^{\rm c}$
as the sets 
\begin{equation}
d_1^{\rm c} = \{ (\omega , k) \in \mathbb{C}^2  \, | \, k^4 - \sigma \omega^2 = 0 \},
\qquad 
d_2^{\rm c} = \{ (\omega , k) \in \mathbb{C}^2 \,  | \, k^2 - \omega^2 / c^2 = 0 \}. 
\label{eq1011a}
\end{equation}
Indeed, the real dispersion diagrams are subsets of corresponding complex dispersion diagrams.
Analytical continuation of a dispersion diagram proved itself to be a useful tool 
in the analysis of transient processes in waveguides  
\cite{Randles1969,Randles1971,Shanin2017,Shanin2018a}. 

The points of $d_1^{\rm c}$ correspond to the  
waveguide modes in the plate, namely to the functions  
\[
\zeta(t, x) = \zeta_0 e^{\i k x - \i \omega t}, 
\]
which 
are solutions of the equation of motion for an unloaded plate without sources. 
The similar statement is valid for the set  
$d_2^{\rm c}$ and the gas layer with Neumann walls. Indeed, $d_2^{\rm c}$
is the dispersion diagram for piston modes in such a layer.

Let $(\omega_* , k_*)$ be a crossing point of the diagrams $d_1^{\rm c}$
and $d_2^{\rm c}$:
\begin{equation}
\omega_* = c^2 \sigma^{1/2}, \qquad k_* = c \sigma^{1/2}.
\label{eq1012d}
\end{equation}
All intersection of the dispersion diagrams $d^{\rm c}_1$ and $d^{\rm c}_2$ are as follows: 
\begin{equation}
d_1^{\rm c} \cap d_2^{\rm c} = \{
(0,0), \, (\omega_* , k_*), \, (-\omega_* , -k_*), \, (\omega_* , -k_*), \,
(-\omega_* , k_*)\}.
\label{eq5001b} 
\end{equation}
Indeed, all these points are real. 

\begin{figure}[ht]
\centerline{\epsfig{file=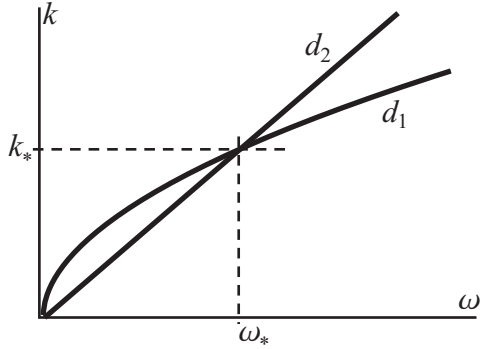}}
\caption{Real dispersion diagrams $d_1$, $d_2$ and the crossing point 
$(\omega_* , k_*)$}
\label{fig02}
\end{figure}

The real dispersion diagrams $d_1$ and $d_2$ and the crossing point 
$(\omega_* , k_*)$ are sketched in Fig.~\ref{fig02}. 

%%%%%%%%%%%%%%%%%%%%%%%%%%%%
%%%%%%%%%%%%%%%%%%%%%%%%%%%%%%%%%%%%%%% 

%%%%%%%%%%%%%%%%%%%%%%%%%%%%%%%%%%%%%%%%%%%%% 
\subsection{Problem 2. 1D bending plate loaded by a 2D half-space}

Let the gas occupy the half-plane $z > 0$, $-\infty <x < \infty$ and the plate (the same as in the previous subsection) is attached to the half-plane along the line 
$z = 0$ (see Fig.~\ref{fig01}, right). 
The wave process in the gas is described by the wave equation Eq.~(\ref{eq1002}), 
and the plate is described by equation Eq.~(\ref{eq1003}). Condition Eq.~(\ref{eq1005a})
is omitted, and Eq.~(\ref{eq1005}) remains valid. 
It is not necessary to introduce a radiation condition, since for a causal solution it should be valid automatically. 

Let us look for the acoustic pressure in the gas near the plate, i.~e.\ at $z=0$.
The solution of the problem can be obtained using the same method as above:
\begin{equation}
p(t,x,0) = - \frac{\rho f_0}{4 \pi^2}
\int \limits_{-\infty}^{\infty} 
\int \limits_{-\infty + \i \epsilon}^{\infty + \i \epsilon}
\frac{
\omega^2 \exp\{\i k x - \i \omega t\}
}{
i T(k^4 - \sigma \omega^2)  \gamma(\omega, k)  + \rho \omega^2 
} 
\rmd \omega \, \rmd k .
\label{eq1013}
\end{equation}    
The air loading is assumed to be light, thus representation Eq.~(\ref{eq1013}) 
can be simplified as follows: 
\begin{equation}
p(t,x,0) = \frac{\i\rho f_0}{4 \pi^2 T}
\int \limits_{-\infty}^{\infty} 
\int \limits_{-\infty + \i \epsilon}^{\infty + \i \epsilon}
\frac{
\omega^2 \exp\{\i k x - \i \omega t\}
}{
(k^4 - \sigma \omega^2) \, \gamma (\omega , k)
} 
\rmd \omega \, \rmd k.
\label{eq1014}
\end{equation}    
The zeros of the denominator of Eq.~(\ref{eq1014}) are described by the 
same functions $d_1^{\rm c}$ and $d_2^{\rm c}$.

%%%%%%%%%%%%%%%%%%%

\subsection{Problem 3. Ice, air, and water in 3D space}

%\redtext{The whole subsection is added in the revised version}

The previous two problems are relatively simple, and they will  
be studied in details below. In this subsection 
we formulate mathematically the problem that motivated our research, namely 
the 3D problem of excitation of an ice plate loaded by a water substrate below 
and by a light air above. The excitation is a pulse of force applied to the 
ice plate at some point. The receiver is located in the air near the ice at the distance 
$L$ from the source point.

This problem is considerably more complicated than the model ones, so we are not planning 
to study it in details. However, we are going to describe the term related to the 
coincidence point using the method developed on the basis of the model problems. 
In this case, the coincidence point occurs as a crossing point of the
dispersion diagram of the air and the dispersion diagram of the bending waves in the ice 
on the water substrate.     

\begin{figure}[ht]
\centerline{\epsfig{file=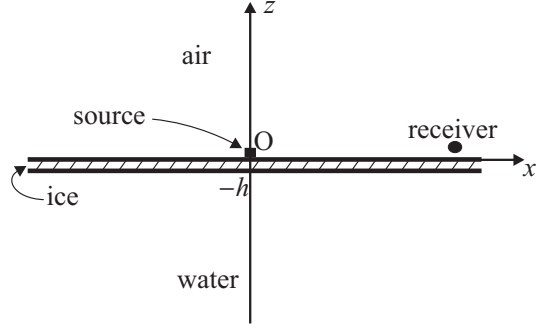, width= 7cm}}
\caption{Geometry of Problem~3. Axis $y$ is directed normally to the plane 
of the figure}
\label{fig01a}
\end{figure}

The geometry of the problem is shown in Fig.~\ref{fig01a}. The $y$-axis is directed 
normally to the plane of the figure and is not shown. The water occupies the domain 
$z < -h$, all the rest is similar to Problem~2.   

%%%%%%%%%%%%%%%%%%%%%%%%%%%%
%%%%%%%%%%%%%%%%%%%%%%%%%%%%%%%%%%%%%%%  

The acoustic field in the air is described by the wave equation 
\begin{equation}
\left(
\frac{\ptl^2}{\ptl x^2}
+ 
\frac{\ptl^2}{\ptl y^2}
+ 
\frac{\ptl^2}{\ptl z^2}
-
\frac{1}{c^2}  
\frac{\ptl^2}{\ptl t^2}
\right) \phi (t , x, y, z) = 0
\label{eq1002a}
\end{equation}
($c$ is the wave velocity in the air).
The acoustic field in the water is described by the wave equation 
\begin{equation}
\left(
\frac{\ptl^2}{\ptl x^2}
+ 
\frac{\ptl^2}{\ptl y^2}
+ 
\frac{\ptl^2}{\ptl z^2}
-
\frac{1}{c_w^2}  
\frac{\ptl^2}{\ptl t^2}
\right) \phi_w (t , x, y, z) = 0.
\label{eq1002c}
\end{equation}
Here $c_w$ is the wave velocity in the water, and $\phi_w$ is the acoustic potential in the water. The acoustic pressure in the water 
is 
\begin{equation}
p_w = \rho_w \frac{\ptl \phi_w}{\ptl t},
\label{eq1001a}
\end{equation}
where $\rho_w$ is the density of the water. The pressure in the air is 
defined by the first equation of Eq.~(\ref{eq1001}).

We assume that the ice bears only the bending waves described by the equation
\[
\left[ 
T \left( 
\frac{\ptl^4}{\ptl x^4}
+ 2 \frac{\ptl^4}{\ptl x^2 \, \ptl y^2}
+ \frac{\ptl^4}{\ptl y^4}
\right) 
+ \rho_p h \frac{\ptl^2}{\ptl t^2}
\right] \zeta (t, x, y)
\] 
\begin{equation} 
+ p(t,x,y,0)  - 
p_w(t,x,y,-h) = -f_0 \,\delta(t) \delta(x) \delta(y),  
\label{eq1003a}
\end{equation} 
which is a modification of Eq.~(\ref{eq1003}). $T$ is the flexural stiffness of the ice (see 
Eq.~(\ref{eq1004})), and $\rho_p$ is the density of the ice. 

The continuity condition for the air / ice interface is Eq.~(\ref{eq1005}), 
and for the water / ice interface is 
\begin{equation}
\left.
\frac{\ptl \phi_w}{\ptl z} \right|_{z = -h}
= - \frac{\ptl \zeta}{\ptl t}.
\label{eq1005b}
\end{equation}

We have to find the air pressure $p(t , L, 0 , 0)$. 

The system of equations formulated above can be easily solved by applying the  Laplace transform with respect to $t$ and the Fourier transform with respect to $x$ and $y$. The result can be written in the form of the 
Fourier--Bessel integral: 
\[
p (t , L, 0 , 0)  = 
\]
\begin{equation}
\frac{ \i \rho f_0}{4 \pi^2}
\int \limits_{-\infty + \i \epsilon}^{\infty + \i \epsilon}
\int \limits_0^\infty
\frac{\omega^2 k J_0 (kL) \exp \{ - \i \omega t \} }{
\gamma(\omega , k) [T(k^4 - \sigma \omega^2) - i \gamma_w^{-1}(\omega, k) \rho_w \omega^2 ]
- \i \rho \omega^2 
} \rmd k \, \rmd \omega, 
\label{eq1013a}
\end{equation}
where 
\begin{equation}
\gamma_w(\omega, k) \equiv \sqrt{\omega^2 / c_w^2- k^2},
\label{eq1009a}
\end{equation}
and $J_0$ is the Bessel function.

Introduce the dispersion equation for bending waves in ice loaded by water: 
\begin{equation}
D(\omega , k) \equiv (k^4 - \sigma \omega^2) - \frac{\i \rho_w \omega^2}{T \gamma_w (\omega, k)}. 
\label{eqF01}
\end{equation}

In our consideration we assume that the air loading is light, and water loading is 
considerable. Thus, at least locally, one can neglect the last term in the denominator and 
obtain a simplified representation for the pressure:
\begin{equation}
p (t , L, 0 , 0)  = 
\frac{ \i \rho f_0 }{4 \pi^2 T}
\int \limits_{-\infty + i \epsilon}^{\infty + i \epsilon}
\int \limits_0^\infty
\frac{\omega^2 k J_0 (kL) \exp \{ - \i \omega t \} }{
\gamma(\omega , k) D(\omega , k)
} \rmd k \, \rmd \omega, 
\label{eq1014a}
\end{equation} 

One can see that there are two factors in the denominator
of the integrand. One corresponds to 
waves in the air, while the other corresponds to waves in the ice plate loaded 
by the water. The coincidence point $(\omega_*^\dag , k_*^\dag)$ is the common zero of 
both functions: 
\begin{equation}
(\omega_*^\dag)^2 / c^2 - (k_*^\dag)^2 = 0 , 
\qquad 
D(\omega_*^\dag ,  k_*^\dag) = 0.
\label{eqF02}
\end{equation}
Our aim is to describe the processes described by the integral Eq.~(\ref{eq1014a})
estimated near this point. 
\color{black}

%%%%%%%%%%%%%%%%%%%%%%%%%%%%%%%%%%%%%%%%%%%%%%%

\section{Computation of the integral Eq.~(\ref{eq1012}) (Problem~1)} 

The integral Eq.~(\ref{eq1012}) can be easily computed analytically. 

For any $\omega$ with ${\rm Im}[\omega] = \epsilon$,
the integrand is a regular function on the real axis of $k$ and near 
this real axis. For simplicity, 
slightly deform the integration contour  
in the $k$-plane 
such that the deformed contour
coincides with the real axis almost everywhere, 
and bypasses {\em above\/} the points $0, \pm k_*$. 
Denote this contour by~$R_+$.
This choice of the contour
is arbitrary, and the result remains the same if the 
contour passes below the points $0, \pm k_*$. 

First, take the internal integral in the $\omega$-domain using the 
residue method. The contour of integration should be closed in the 
lower half-plane, since the factor $e^{- \i \omega t}$ decays there 
for $t > 0$. For a fixed $k$, there are four poles 
$\omega = \pm k c$, $\omega = \pm \sigma^{-1/2} k^2$, 
and all of them fall within the closed contour. The result is a 
\begin{equation}
p(t, x, 0) = I_1 (t, x) + I_2 (t, x) + I_3 (t, x) + I_4 (t, x),
\label{eq8001}
\end{equation}
where 
\begin{equation}
I_{1,2}(t,x) = \mp \frac{\i \rho f_0 c^2}{4 \pi  T H}
\int_{R_+}
\frac{
\exp \{ \i k (x \mp  c t ) \}
}{
k (k^2 - k_*^2)
}
\rmd k,
\label{eq8002}
\end{equation}
\begin{equation}
I_{3,4} (t, x) = \pm \frac{\i \rho f_0 c^2}{4 \pi T H}
\int_{R_+}
\frac{
\exp \{ \i k x \mp \i \sigma^{-1/2} k^2 t \}
}{
k^2 - k_*^2
}
\rmd k.
\label{eq8003}
\end{equation}
The integrals $I_{1,2}$ can be taken using the residue method: 
\begin{equation}
I_{1,2} (t, x) = 0 , \qquad x > \pm ct,
\label{eq8004}
\end{equation}
\begin{equation}
I_{1,2} (t, x) = \pm
\frac{\rho f_0}{4 T H \sigma} 
\left( 
2 - \exp \{ \i k_* x \mp i \omega_* t\} -  \exp \{ - \i k_* x \pm i \omega_* t\} 
\right), 
\quad 
x < \pm c t. 
\label{eq8005}
\end{equation}  

The integrals $I_{3,4}$ can be expressed through the Fresnel's integrals: 
\[
I_3 (t, x) =
\]
\begin{equation}
 \frac{\rho f_0}{4 T H \sigma}
\left( 
e^{ \i k_* x - \i \omega_* t}
C \left( \frac{\sigma^{1/4} (x - 2 c t)}{  2t^{1/2}} \right) 
-
e^{ -\i k_* x - \i \omega_* t}
C \left( \frac{\sigma^{1/4} (x + 2 c t)}{ 2 t^{1/2} }\right)
\right), 
\label{eq8006}
\end{equation}
\[
I_4 (t, x) = 
\]
\begin{equation}
\frac{\rho f_0}{4 T H \sigma}
\left( 
e^{ -\i k_* x + \i \omega_* t }
\bar C \left( \frac{\sigma^{1/4} (x - 2 c t)}{  2t^{1/2}} \right) 
-
e^{ \i k_* x  + \i \omega_* t }
\bar C \left( \frac{2\sigma^{1/4} (x + 2 c t)}{  2t^{1/2}} \right)
\right), 
\label{eq8007}
\end{equation}
where $C(a)$ for real $a$ is the Fourier integral 
defined by Eq.~(\ref{eq2029}) in Appendix~A3. Representations Eq.~(\ref{eq8006}) and Eq.~(\ref{eq8007})
follow from the non-trivial formula  Eq.~(\ref{A213}). 
The bar sign denotes the complex conjugation here and below.

Let us analyze the solution.
Introduce the ``formal velocity''
\begin{equation}
V \equiv \frac{x}{t}.
\label{eq8008}
\end{equation}
Consider the asymptotics $x \to \infty$, $V = \mbox{const}$.
Introduce  also the value 
\begin{equation}
\eta = \frac{c^{1/2}}{x^{1/2} \sigma^{1/4}}
\label{eq5019}
\end{equation}
and the function 
\begin{equation}
F_{\pm}(x, V) = \frac{ V^{1/2} }{4\sqrt{\pi} x^{1/2} \sigma^{1/4} (V \mp 2c)}
\exp \left\{ 
\i\frac{(V \mp 2c)^2 \sigma^{1/2} x}{ 4V }
+ \i\frac{\pi }{4} \right\} 
\label{eq8009}
\end{equation}

Use the asymptotics Eq.~(\ref{eq2029z}) and Eq.~(\ref{eq2029y})
for $C(\xi)$ for $|\xi| \gg1$.
As the result, if $|V - 2 c| \gg \eta$ the asymptotics is as follows: 
\begin{equation}
p(t,x, 0) \approx \frac{\rho f_0 }{4 H  T \sigma} 
(u_1(x, t) + u_2(x, t) + u_3(x, t)) + \mbox{C.C.}
\label{eq8009a}
\end{equation}
where C.C.\ denotes the complex conjugated terms, 
$u_1 (x, t)$ is the pulse mainly attributed to the plate wave: 
\begin{equation}
u_1(x, t) =
e^{  \i k_* x - \i \omega_* t } F_+(x , x/t )  - 
e^{ -\i k_* x - \i \omega_* t } F_-(x , x/t ) ,
\label{eq8010}
\end{equation}
$u_2 (x, t)$ is the pulse mainly attributed to the piston wave in the gas waveguide: 
\begin{equation}
u_2(x, t) =
\left\{ \begin{array}{ll} 
1,                                & 0 < x/t < c, \\
0,                                & x/t > c ,
\end{array} \right.
\label{eq8011}
\end{equation}
and $u_3$ is the 
\redtext{pulse related to the coincidence point of the dispersion diagram}:
\begin{equation}
u_3(x, t) =
\left\{ \begin{array}{ll} 
\exp\{ \i k_* x - \i \omega_* t \}, & c < x/t < 2 c, \\
0        ,                        & 0 < x/t < c \quad \mbox{or} \quad x /t > 2c,
\end{array} \right.
\label{eq8012}
\end{equation}
Note that $p(t, -x, 0) = p(t, x, 0)$ due to the problem symmetry.

In the intermediate zone $|V - 2 c| \sim \eta$ the argument of function $C$ is of order of~1.
No asymptotics of $C$ can be used in this zone, so one should use the function $C$
itself.
For a fixed large $x$, the width of the intermediate zone in the $t$ variable is as follows: 
\begin{equation}
\Delta t \sim \frac{x^{1/2}}{c^{3/2} \sigma^{1/4}}.
\label{eq8012a}
\end{equation} 
Thus, this width grows as $x^{1/2}$. It is important that this width grows, but slower 
than linearly.  

The \redtext{term $u_3$} is a purely monochromatic pulse with a smooth front at $V \approx 2c$ 
and an abrupt front at $V = c$. 
The values $V= c$ and $V = 2c$ are, thus, the boundaries of the domain occupied 
by the \redtext{term $u_3$} in the $(x, V)$  plane. These values emerged in our research 
quite naturally in the process of getting the explicit solution. We should note that 
$2 c$ and $c$
are the values of the group velocity of the branches of the dispersion diagram at the 
crossing point $(\omega_* , k_*)$. Below we discuss this feature in details and demonstrate that the 
\redtext{
pulse related to the coincidence point of the dispersion diagram} 
always have fronts 
linked to 
the group velocities 
of the interacting modes. 

Note also that the field is continuous at $x = ct$ since the sum $u_2 + u_3$ is continuous 
there.

%%%%%%%%%%%%%%%%%%%%%%%%%%%%%%%%%%%%%%%%%%%%%%%%%%%%%%%%%%%%%%%%%%

\section{Estimation of Eq.~(\ref{eq1014}) (Problem~2)}

Unfortunately, the integral Eq.~(\ref{eq1014}) cannot be taken explicitly. So our next aim is to develop a technique of evaluation of such an integral in a general situation. 
As above, we fix the value $V = x / t$ and take $x \to \infty$
to build the asymptotic procedure.   
The idea of the technique is quite standard: the surface of integration should be deformed
in such a way that the integrand is exponentially small everywhere except neighborhoods 
of several ``special points''. These points are the 
saddle points on the real dispersion diagrams and the crossing points of the dispersion diagrams. The integrals over the neighborhoods of the ``special points'' can be taken 
approximately (asymptotically). Some typical integrals of this sort are listed 
in the Appendix.

\subsection{Overview of the surface deformation procedure. The ``main statement'' and its proof}

\label{sec_main}

%Consider an integral of a general form 
%\begin{equation}
%u (x, V) = 
%\int \limits_{-\infty}^{\infty}
%\int \limits_{-\infty + i \epsilon}^{\infty + i \epsilon}
%\frac{A(\omega , k)
%\exp \{ i x (k - \omega /V)\}
%}{
%D_1(\omega ,k) D_2(\omega,k)} 
%\rmd \omega \, \rmd k, \qquad V \equiv x/ t.
%\label{B001}
%\end{equation}
%Function $A$ is assumed to be holomorphic everywhere. 
%Functions $D_1$ are $D_2$ are dispersion functions
%that have zero sets $d^{\rm c}_1$, $d^{\rm c}_2$ in $\mathbb{C}^2$.
%The zero sets can be also branch sets (here we do not consider the case of branch 
%set different from the zero set, although this case is important for applications).
%The sets $d^{\rm c}_{1,2}$ are singularity sets of the integrand.
%The real subsets of $d^{\rm c}_{1,2}$ are denoted by $d_{1,2}$. They are
%referred to as the real branches of the dispersion diagram.   

At each point of the branches $d_j^{\rm c}$ of the dispersion diagrams 
one can define the group velocity 
\begin{equation}
v_{\rm gr} = \frac{\rmd \omega}{\rmd k}.
\label{B001a}
\end{equation} 
The group velocities at the points of the real branches $d_j$ are real. 

The saddle points on the real branches of the dispersion 
diagrams are the points at which 
\begin{equation}
v_{\rm gr} = V.
\label{B001aa}
\end{equation} 
Let us find the position of the saddle point on $d_1$, for which 
$\omega = \sigma^{-1/2} k^2$. The group velocity 
is then
\begin{equation}
v_{\rm gr} = 2 \sigma^{- 1/2} k .
\label{B001z}
\end{equation}
The saddle point has coordinates $(\omega_{\rm s} , k_{\rm s})$ with 
\begin{equation}
\omega_{\rm s} =  \omega_{\rm s} (V) =\frac{\sigma^{1/2} V^2}{4},
\qquad 
k_{\rm s} = k_{\rm s}(V) = \frac{\sigma^{1/2} V}{2} .
\label{B001y} 
\end{equation}
 
The main statement of the 
estimation procedure is as follows: 

{\em For almost all values of $V$, the terms of the field  not 
decaying exponentially as 
$x \to \infty$ are produced by the fragments of the integration surface that are 
located in neighborhoods of either the saddle points on the real branches, or the crossing points of the real branches of the dispersion diagrams.}

The idea of the proof is as follows. Consider the integral Eq.~(\ref{eq1014}) as an integral of 
an
analytic {\em differential 2-form} over some surface (smooth manifold) \cite{Shabat1992}:
\begin{equation}
u(x , V) = \int_{\Gamma_0} \Psi,
\label{B002}
\end{equation}
where  
\begin{equation}
\Psi = \frac{\exp \{ \i x (k - \omega /V)\}}{(k^4 - \sigma \omega^2) \gamma(\omega , k)} 
\, 
\rmd \omega \wedge \rmd k,
\label{B003}
\end{equation}
and $\Gamma_0$ is an oriented manifold with ${\rm Im}[k] = 0$, ${\rm Im}[\omega] = \epsilon$.
According to the 2D Cauchy's theorem, one can deform continuously the manifold $\Gamma_0$, 
and the value of the integral should remain the same if the manifold
does not cross the singular sets $d_j^{\rm c}$
of the integrand form  
during the course of deformation. 
  
To estimate the integral,   
one has to find a deformation of $\Gamma_0$ 
into a new manifold $\Gamma_1$, such that
the integrand is exponentially small  {\em almost\/} everywhere, i.~e.\  such that
\begin{equation}
x \, \left( {\rm Im}[k] - {\rm Im}[\omega] / V \right) \gg 1.
\label{B004}
\end{equation}
The neighborhoods on $\Gamma_1$ where one {\em cannot\/} fulfill the inequality
Eq.~(\ref{B004}) are used to build the estimation of the wave field components. 

We consider only small deformations of $\Gamma_0$. 
Since $x$ is large,  
a small deformation is enough to fulfill Eq.~(\ref{B004}). 

Let the deformed integration manifold $\Gamma_1$ be 
parametrized by real parameters $(\omega', k')$ as follows: 
\begin{equation}
\Gamma_1 : 
\qquad  
\omega = \omega' + \i f_1 (\omega', k'),
\quad   
k = k' + \i f_2 (\omega', k'),
\label{B005}
\end{equation}   
where $f_1$, $f_2$ are some smooth real functions. 
The deformation process can be described using a parameter $\chi \in [0,1]$. 
For this, introduce a family of integration surfaces:
\begin{equation}
\Gamma (\chi) : 
\qquad  
\omega = \omega' + \i \chi f_1 (\omega', k') + \i(1-\chi) \epsilon,
\quad   
k = k' + \i \chi f_2 (\omega', k').
\label{B006}
\end{equation}   
One can see that 
\[
\Gamma(0) = \Gamma_0, \qquad \Gamma(1) = \Gamma_1 .
\] 

Let $(\omega_0 ,k_0)$ be a real point belonging to the dispersion diagram~$d_j$.
Consider a small complex neighborhood of $(\omega_0 , k_0)$. Let us describe the eligible 
deformations of $\Gamma_0$ in this neighborhood, i.~e.\ the deformations in which the 
manifold of integration does not cross~$d_j^{\rm c}$. 

Let the group velocity of the corresponding branch of the dispersion 
diagram at $(\omega_0 , k_0)$ be equal to $v$.
A point $(\omega , k)$ in this neighborhood belongs to the complex branch of 
the dispersion diagram only if 
\begin{equation}
{\rm Im}[k] \approx \frac{{\rm Im}[\omega] }{ v}
\label{B007}
\end{equation}
(indeed, this is not a sufficient condition). This follows from the fact that 
\begin{equation}
k - k_0 \approx \frac{\omega - \omega_0 }{ v},
\label{B008}
\end{equation}
and Eq.~(\ref{B007}) is the imaginary part of Eq.~(\ref{B008}).

\begin{figure}[ht]
\centerline{\epsfig{file=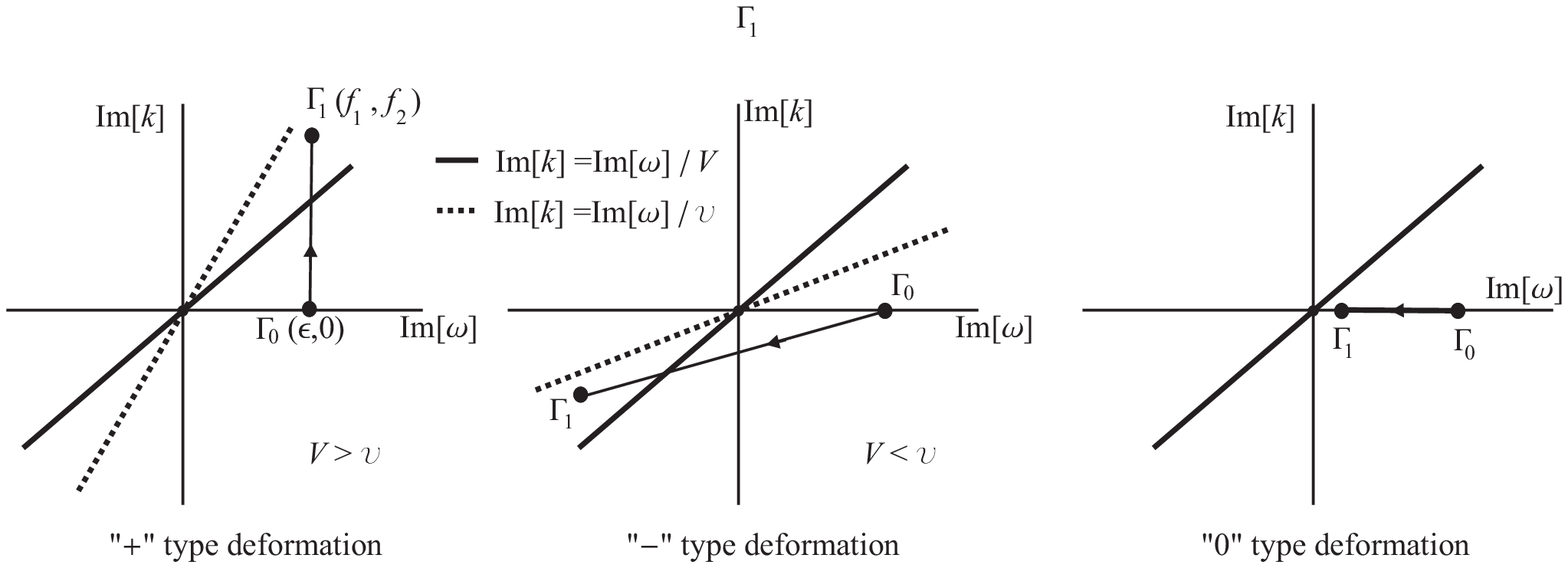, width = 17cm}}
\caption{Types of deformation of $\Gamma_0$}
\label{fig16}
\end{figure}

Let the deformation 
of $\Gamma_0$ in the neighborhood of $(\omega_0 , k_0)$ be chosen in such 
a way that 
the functions $f_1$, $f_2$ are locally constant.
Let be $V > v$.   
Consider the conditions Eq.~(\ref{B007}) and Eq.~(\ref{B004}) graphically. 
It is a difficult task to display a 2D manifold in a 4D space, so 
we choose to plot projections of the integration manifolds onto the  
coordinate plane $({\rm Im}[\omega] , {\rm Im}[k])$. A fragment 
of the initial manifold $\Gamma_0$ is 
projected onto a single 
point $(\epsilon, 0)$ in this plane. Corresponding fragment of the deformed 
manifold $\Gamma_1$ is shown also as a single point 
$(f_1 (\omega_0 , k_0) , f_2 (\omega_0 , k_0))$. The deformation process 
is the motion along the segment connecting these points (see Fig.~\ref{fig16}, left).  

%%%%%%%%%%%%%%%%%%%%%%%%%%%%
%%%%%%%%%%%%%%%%%%%%%%%%%%%%%%%%%%%%%%% 

The inequality Eq.~(\ref{B004}) is fulfilled is the resulting point $(f_1, f_2)$ 
is located above the 
line ${\rm Im} [k] = {\rm Im} [\omega] / V$ (with a margin of the height
equal to $1/x$). The line 
${\rm Im} [k] = {\rm Im} [\omega] / V$
is shown by the bold solid line in the figure. 

The manifold $\Gamma(\tau)$ can hit 
the singularity {\em only if\/} the segment $[\Gamma_0, \Gamma_1]$
crosses the line on which Eq.~(\ref{B007}) is valid. This
line is shown dotted in the figure. 
This means that if the segment $[\Gamma_0 , \Gamma_1]$ does not cross the 
dotted line, then the deformation is eligible. 

An inverse statement is much more subtle, but it also can be proven: If the
segment  $[\Gamma_0, \Gamma_1]$ crosses the dotted line in the diagram, then 
$\Gamma(\tau)$ crosses the singularity set at some points, and the deformation is not eligible.

One can see that the eligible deformation in the case $V > v$ looks like it is shown in 
Fig.~\ref{fig16}, left. The point $\Gamma_1$ belongs to the first quadrant of the 
coordinate plane.
This deformation will be referred to as a deformation of the ``$+$''~type. 

If $V < v$ then the eligible deformation is as shown in Fig.~\ref{fig16}, center.
The point $\Gamma_1$ now belongs to the third quadrant of the coordinate plane. 
This deformation will be referred to as the deformation of the ``$-$''~type.

For the ``special points'' (the saddle points or the crossing points 
of the dispersion diagrams)
we need a deformation that does not obey Eq.~(\ref{B004}), but still 
does not cross the singularities. Such a deformation is called ``0''~type, and is shown
in Fig.~\ref{fig16}, right.  
   
Let $v$ grow continuously 
from $v < V$ to $v > V$. The point $\Gamma_1$ then goes far into the first quadrant.
One can see that the ``$+$''~type deformation cannot be transformed continuously into the ``$-$''~type deformation. Therefore, between the domains of ``$+$''~type deformation 
and ``$-$''~type deformation there should be 
a zone with ``$0$''~type deformation. This is a very important conclusion, since 
the domains with ``$+$'' or ``$-$''~type deformation do not produce non-vanishing 
field components, while the domains with the ``0''~type can produce such components. 

\begin{figure}[ht]
\centerline{\epsfig{file=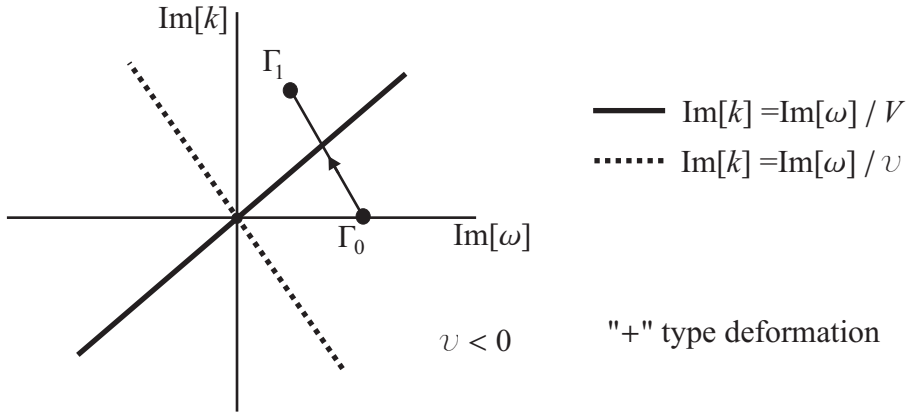}}
\caption{Deformation of $\Gamma_0$ when $v < 0$}
\label{fig17}
\end{figure}

Finally, if $v < 0$, only a ``$+$''~type deformation is eligible, and it looks as shown in Fig.~\ref{fig17}.   

%%%%%%%%%%%%%%%%%%%%%%%%%%%%
%%%%%%%%%%%%%%%%%%%%%%%%%%%%%%%%%%%%%%% 
 
The deformation should be made carefully only near the real branches of the dispersion diagram.
We assume that an eligible smooth deformation can be easily found
in the rest of the $(\omega', k')$-plane, since there are no obstacles for deformation at such places.
This is why, below we indicate the type of the deformation only for the neighborhoods of the 
real branches of the dispersion diagram.

%%%%%%%%%%%%%%%%%%%%%%%%%%%%%%%%%%%%%%%%%%%%%%%%%%%%%%%%%%%%

\subsection{Integration surface deformation for the integral Eq.~(\ref{eq1014})}

In this subsection we analyze the deformation of the integration surface for 
Eq.~(\ref{eq1014}) in terms of the types of the deformation 
introduced above. 
Note that the singular sets of Eq.~(\ref{eq1012}) and Eq.~(\ref{eq1014})
are the same, so the conclusions obtained here can be applied to Eq.~(\ref{eq1012}) as well, 
and the exact value of Eq.~(\ref{eq1012}) can be examined on its compliance with our asymtotic theory.  

The integrand of Eq.~(\ref{eq1014}) has 
the real branches of the dispersion diagram named $d_1$ and~$d_2$.
The group velocity takes all values from $-\infty$ to $\infty$ on $d_1$.
For each $V \ne 0$, there exist two  saddle points on~$d_1$. 
On $d_2$, the group velocity takes values~$\pm c$ only. 

There are five crossing points Eq.~(\ref{eq5001b}) 
of the dispersion diagrams $d_1$ and~$d_2$. The 
group velocity on $d_1$ is equal to $\pm 2c$ 
at the non-zero crossing points.   
  
\begin{figure}[ht]
\centerline{
\epsfig{file=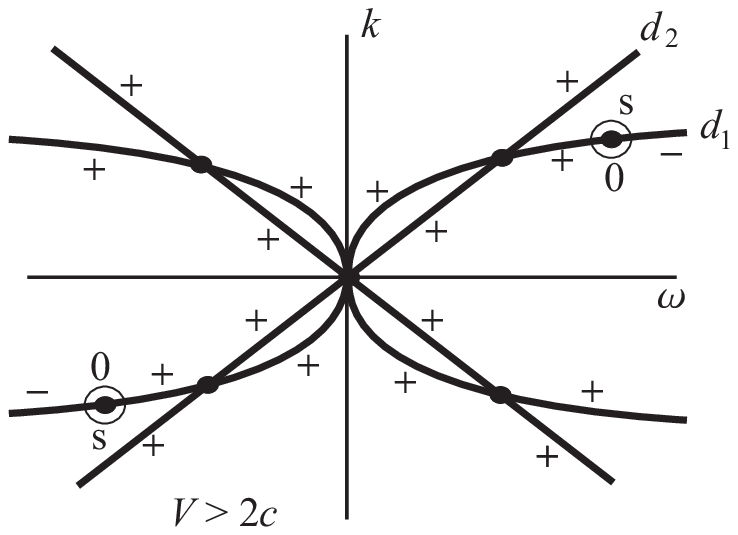}
\qquad \qquad
\epsfig{file=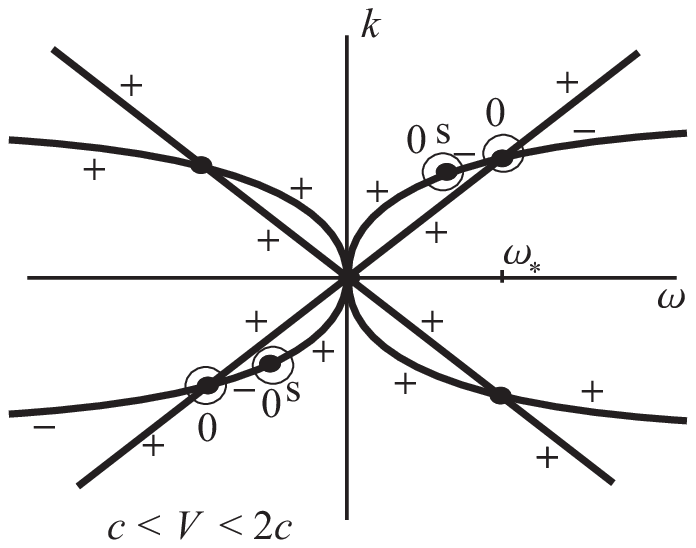}
}
\caption{Deformation diagram for $V > 2 c$ (left), and for 
$c < V < 2 c$ (right)
}
\label{fig18}
\end{figure}

Let be $V > 2 c$. Fig.~\ref{fig18}, left,  shows how the 
deformation type should be chosen in this case.
The saddle points on $d_1$ are marked by symbol~``s''.
The ``$+$'', ``$-$'', and ``0''~types of surface deformation 
are marked by corresponding signs.

%%%%%%%%%%%%%%%%%%%%%%%%%%%%
%%%%%%%%%%%%%%%%%%%%%%%%%%%%%%%%%%%%%%% 
 
Consider the saddle point with positive $\omega$ and~$k$.
The group velocity on $d_1$ is bigger than $V$ to the right 
of the saddle point and is smaller than $V$ to the left of the saddle point. 
Thus, near $d_1$, one should choose the ``$-$''~type of the surface deformation to the 
right, and ``$+$''~type of the deformation to the left. Since the deformation 
should be continuous, one should choose the ``0''~type of the deformation 
in the neighborhood of the 
saddle point. 
Thus, the saddle points on $d_1$ produce non-vanishing components of the wave field. 

Indeed, there should be some continuous transition between ``0''~type and the
``$+$''/``$-$''~types.   

All other branches are labeled with ``+''~type deformation.  
In particular, for each of five crossing points, all branches crossing at them
are of the ``$+$''~type. Therefore, the vicinities of the crossing points can be shifted according to the ``$+$''~type, and the crossing points do not produce non-vanishing field components. 

Let be $c < V < 2c$. Corresponding deformation diagram 
is shown in Fig.~\ref{fig18}, right. 
The neighborhood of the crossing point $(\omega_* , k_* )$ changes its status 
comparatively to Fig.~\ref{fig18}, left. The 
branch $d_1$ should be deformed according to the ``$-$''~type, while $d_2$ 
should be deformed according to the ``$+$''~type. 
Since the deformation is continuous, the neighborhood 
of the crossing point should be of the ``$0$''~type. Therefore, the the crossing point 
produces a non-vanishing field component. 
Besides, the  saddle point on $d_1$ 
(also deformed according to the ``$0$''~type)
produces another field component. 

%%%%%%%%%%%%%%%%%%%%%%%%%%%%%
%\begin{figure}[ht]
%\centerline{\epsfig{file=Figures/Fig19.eps}}
%\caption{Deformation diagram for $c < V < 2 c$}
%\label{fig19}
%\end{figure}
%%%%%%%%%%%%%%%%%%%%%%%%%%%%%%%%%%%%%%% 

\begin{figure}[ht]
\centerline{\epsfig{file=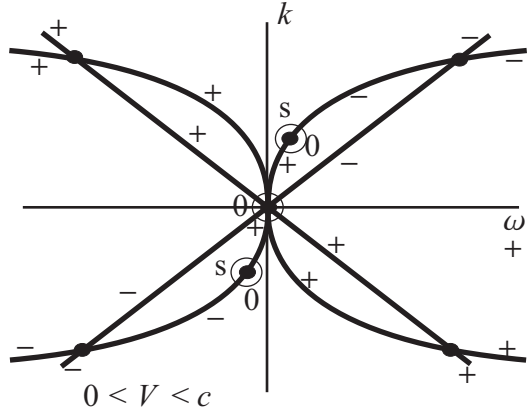}}
\caption{Deformation diagram for $0 < V <  c$}
\label{fig20}
\end{figure}

Consider the case $0< V < c$. The diagram is shown in Fig.~\ref{fig20}. 
The neighborhood of the crossing  point $(\omega_* , k_*)$ can now be deformed according to 
the ``$-$''~type, and thus it does not produce non-vanishing field components.  
The crossing point $(0,0)$ should be deformed according to the 
``$0$''~type, since one branch crossing at this point is deformed according to 
the ``$-$''~type, while all other branches are deformed according to the 
``$+$''~type. 
Besides, there are saddle points on $d_1$ that also produce non-vanishing components.  

%%%%%%%%%%%%%%%%%%%%%%%%%%%%
%%%%%%%%%%%%%%%%%%%%%%%%%%%%%%%%%%%%%%% 

\begin{figure}[ht]
\centerline{
\epsfig{file=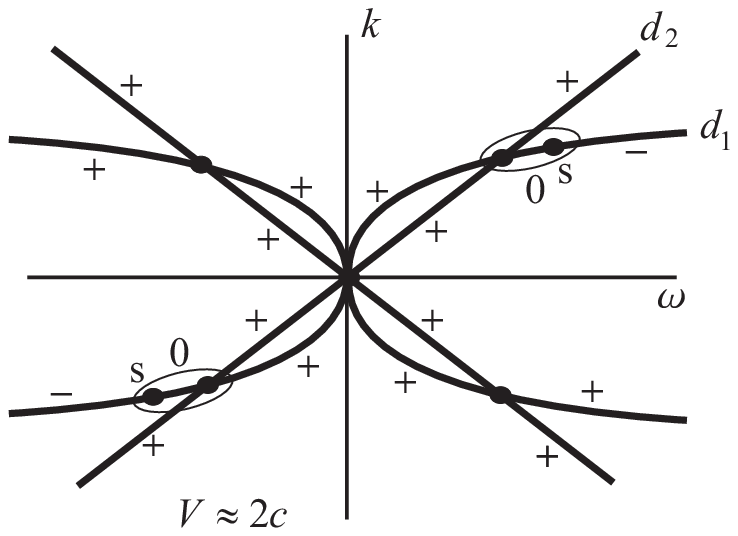}
\qquad \qquad 
\epsfig{file=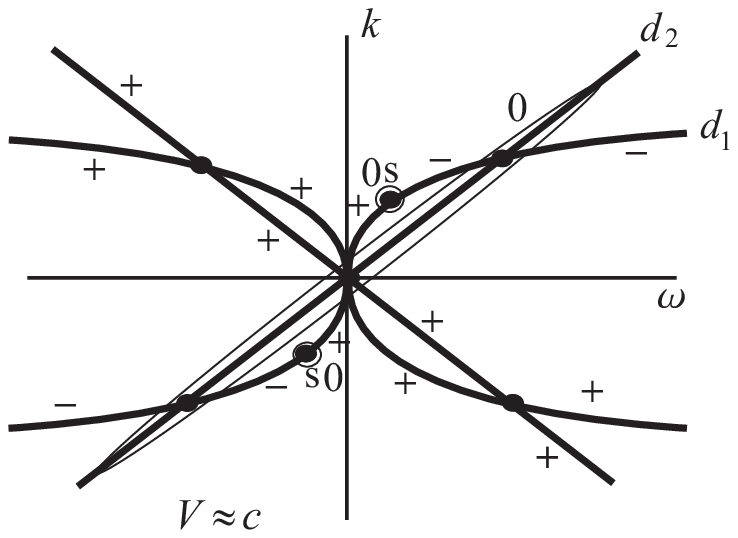}
}
\caption{Deformation diagram for $ V \approx 2 c$ (left), and for $V \approx c$ (right) }
\label{fig18a}
\end{figure}

When $V \approx 2 c$, the saddle point is close to the crossing point 
of the branches of 
the dispersion diagram. 
This case produces an intermediate asymptotics and 
it should be considered in a special way. The diagram is shown 
in Fig.~\ref{fig18a}, left. One can see that the zone of ``0''~type deformation 
covers both the crossing point and the saddle point   

%%%%%%%%%%%%%%%%%%%%%%%%%%%%
%%%%%%%%%%%%%%%%%%%%%%%%%%%%%%%%%%%%%%% 

The most sophisticated case is $V \approx c$. In this case, the whole neighborhood 
of the branch $d_2$ is tangential to the line $k = \omega / V$
(see Fig.~\ref{fig18a}, right), and thus it cannot be 
analyzed locally. 

Let us summarize this diagram consideration. 
The saddle point on $d_1$ produces a non-vanishing component for all $V$, 
the crossing point $(0,0)$ produces a non-vanishing component only for $0 < V < c$,
and the crossing point $(\omega_* , k_*)$ produces a non-vanishing component only for 
$c < V < 2c$.

One can see that corresponding wave components are analogous to
$u_1$, $u_2$, and $u_3$ from Eq.~(\ref{eq8009a}) for the 
integral Eq.~(\ref{eq1012}). The last component is the 
\redtext{pulse related to the coincidence point of the dispersion diagram}. 
Thus, the qualitative analysis based on deformation of the integration surface 
is in agreement with the exact solution.

{\bf Remark.}
The case $V / c \approx 0$ requires a special consideration, since the 
``0''-deformation zone is elongated in the vertical direction. However,
this case is not considered in this paper, since it requires more complicated 
standard integrals.

%%%%%%%%%%%%%%%%%%%%%%%%%%%%%%%%%%%%%%%%%%
\subsection{Local estimations of the integral Eq.~(\ref{eq1014})}

According to the analysis of the surface integral deformation,
if $V$ is not close to $2c$, $c$, or $0$,
one can expect to obtain an asymptotic estimation of  Eq.~(\ref{eq1014}) in the 
form 
\begin{equation}
p (t, x, 0) \approx u_1 (t , x) + \bar u_1 (t , x) + 
u_2(t,x) + u_3(t,x) + \bar u_3 (t,x) ,
\label{eq9001}
\end{equation}
where $\bar{}$ is the complex conjugation operator. The terms in the right are the following wave components: 

\begin{itemize}

\item[]
$u_1$ is produced 
by the saddle point on $d_1$ with positive $\omega$ and $k$, 

\item[] 
$\bar u_1$ is produced 
by the saddle point on $d_1$ with negative $\omega$ and $k$, 

\item[] 
$u_2$ is  produced 
by the crossing point of dispersion diagrams at $(0,0)$, 

\item[]
$u_3$ is  produced 
by the crossing point of dispersion diagrams at $(\omega_*,k_*)$,

\item[]
$\bar u_3$ is  produced 
by the crossing point of dispersion diagrams at $(-\omega_*, -k_*)$.

\end{itemize}

The terms $u_1$ and $\bar u_1$ are non-zero for all $V$, 
the term $u_2$ is non-zero for $0<V<c$, the terms 
$u_3$ and $\bar u_3$ are non-zero only for $c < V < 2c$. 

Note that definition of $u_2$ in Eq.~(\ref{eq9001}) is slightly different from that of 
Eq.~(\ref{eq8009a}) (factor 2 is omitted for convenience). 

In the zone $V \approx 2 c$ the 
estimation of the field can be found in the form 
\begin{equation}
p(t, x, 0) = u_{2 c} (t, x) + \bar u_{2 c} (t, x), 
\label{eq9003}
\end{equation}
where $u_{2c}$ is the term produced by the crossing point $(\omega_* , k_* )$ and the neighboring saddle point. 

In the zone $V \approx c$ the estimation of the field is as follows: 
\begin{equation}
p(t, x, 0) = u_{1} (t, x) + \bar u_{1} (t, x) 
+ u_{c} (t,x), 
\label{eq9003a}
\end{equation}
where $u_1 (t,x)$ is the saddle-point term introduced above, and 
$u_c (t,x)$ is the term produced by the elongated zone located along the 
branch $k = \omega/ c$ in Fig.~\ref{fig18a}, right. 

The width of the zone $V\approx 2c$ can be estimated using a standard 
reasoning based on the concept of the ``domain of influence''~\cite{Borovikov1994}. 
Namely, 
$V$ belongs to the intermediate zone if the phase difference 
between the crossing point $(\omega_* , k_*)$ and the saddle point 
$(\omega_{\rm s}, k_{\rm s})$ is of order of~1. This happens if
\begin{equation}
x | (k_* - \omega_* / V) -  (k_{\rm s}(V) - \omega_{\rm s}(V) / V)| \sim 1.
\label{eq9002}
\end{equation}
Using Eq.~(\ref{B001y}), one can get the estimation $|V - 2 c|\sim \eta$, 
where $\eta$ is defined by Eq.~(\ref{eq5019}), so Eq.~(\ref{eq8012a}) is valid. 

To have a consistent set of asymptitotic expansions, 
the term $u_{2c}$ should have asymptotics 
\begin{equation}
u_{2c} \approx u_1 \qquad \mbox{for} \quad V - 2c \gg \eta,
\label{eq9004}
\end{equation}
\begin{equation}
u_{2c} \approx u_1 + u_3 \qquad  \mbox{for} \quad  2c -V \gg \eta.
\label{eq9005}
\end{equation}

Similarly, one can estimate the width of the intermediate zone $V \approx c$.
The condition is as follows: the phase difference between the points $(0, 0)$
and $(\omega_* , k_*)$ should be of order of~1:
\begin{equation}
x |k_* - \omega_* / V| \sim 1, 
\label{eq9005a}
\end{equation} 
resulting in 
\begin{equation}
|V - c| \sim \delta v , \qquad \delta v = \sigma^{-1/2} x^{-1}. 
\label{eq9005b}
\end{equation} 
For consistency, 
function $u_c$ should have the following asymptotics: 
\begin{equation}
u_{c} \approx u_3 + \bar u_3 \qquad \mbox{for} \quad V - c \gg \delta v,
\label{eq9005c}
\end{equation}
\begin{equation}
u_{c} \approx u_2 \qquad \mbox{for} \quad c -V \gg \delta v.
\label{eq9005d}
\end{equation}

\begin{figure}[ht]
\centerline{\epsfig{file=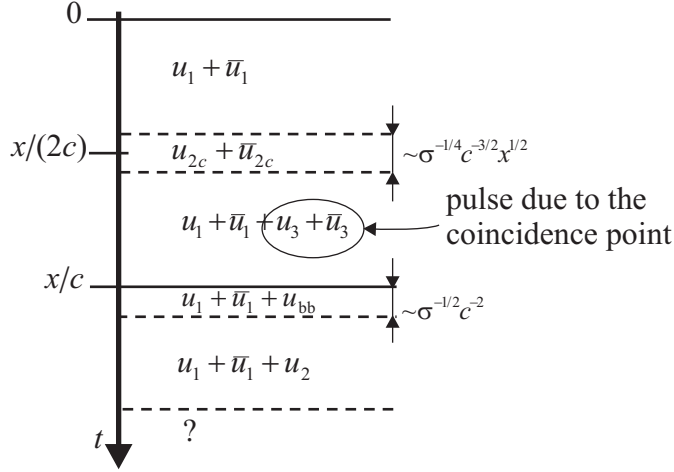}}
\caption{Zones of validity of different asymptotics for a fixed $x$ }
\label{fig21}
\end{figure}

The scheme of all asymptotics is shown in Fig.~\ref{fig21}.  
This scheme shows the zones of validity of asymptotics in the $t$-domain for a large fixed~$x$.
The scheme does not cover small values of $V$ (i.~e.\ large values of~$t$).
The matching rules Eqs~(\ref{eq9004},\ref{eq9005},\ref{eq9005c},\ref{eq9005d}) guarantee 
the consistency of the scheme.

%%%%%%%%%%%%%%%%%%%%%%%%%%%%
%%%%%%%%%%%%%%%%%%%%%%%%%%%%%%%%%%%%%%% 

\redtext{
The terms of Eq.~(\ref{eq9001}) have different physical nature. 
The term $u_1$ is a dispersive bending wave in the plate accompanied by 
some sound pressure in the surrounding air. 
The term $u_3$ is the sound radiated 
into the air
by the supersonic part of the bending pulse. 
The term $u_2$ can be interpreted as the wave traveling in air and radiated 
mainly by the initial impact point.  
}

Below we obtain approximations for the terms of the asymptotics. 

\vskip 6pt
\noindent
{\bf Estimation of $u_1$}
 
Approximate the 
algebraic factors of the 
integrand of Eq.~(\ref{eq1014}) near 
$(\omega_{\rm s} , k_{\rm s})$ as follows:  
\[
\frac{
\omega^2
}{ 
(k^4 - \sigma \omega^2) \sqrt{\omega^2 / c^2 - k^2} 
}
\approx 
\frac{
c
}{
2 \sigma \sqrt{V^2 - 4 c^2}
}
\frac{1}{k -\sigma^{1/4} \omega^{1/2}} .
\]
Evaluation of corresponding integral is shown in Appendix~A2. The parameter 
$\mu$ is equal to~1. The result of application of formula Eq.~(\ref{A105})
is 
\begin{equation}
u_1 (t, x) = \frac{ \rho f_0 c V^{3/2} }{4 \pi^{1/2} T
\sigma^{3/4} x^{1/2} \sqrt{V^2 - 4 c^2} } \exp\{ \i x (k_{\rm s} - \omega_{\rm s} /V)  + 3 \pi \i /4\}.
\label{eq9006}
\end{equation}

\vskip 6pt
\noindent
{\bf Estimation of $u_2$}
 
Approximate the algebraic function in the integrand near the origin
as follows: 
\begin{equation}
\frac{\omega^2}{(k^4 - \sigma \omega^2) \sqrt{\omega^2 /c^2 - k^2}}
\approx \frac{ \i }{\sigma} \frac{1}{ (k - \omega / c)^{1/2}(k + \omega / c)^{1/2}} .
\label{eq9007}
\end{equation}
This approximation can be used if $V$ is not very small comparatively to $c$. 
The integral 
\[
\int \!\!\!\! \int \frac{
\exp \{ \i k x - \i \omega t \} 
}{
(k - \omega / c)^{1/2}(k + \omega / c)^{1/2}
}
\rmd \omega \, \rmd k
\]
can be estimated as the integral Eq.~(\ref{A001}) from Appendix~\ref{AppendixA}. The 
parameters for the integral are 
$\mu_1 = \mu_2 = 1/2$, $\omega_\dag = 0$, $k_\dag  = 0$, 
$v_1 = c$, $v_2 = -c$.
Formula Eq.~(\ref{A014}) can be applied. The result is 
\begin{equation}
u_2 (t,x) = -\frac{\rho f_0 c V}{ \pi T \sigma x \sqrt{c^2 - V^2}  }.
\label{eq9008}
\end{equation} 
  
\vskip 6pt
\noindent
{\bf Estimation of $u_3$}

Estimate the algebraic function in the integrand near the point $(\omega_* , k_*)$
as follows: 
\begin{equation}
\frac{
\omega^2 
}{
(k^4 - \sigma \omega^2) \sqrt{\omega^2 / c^2 - k^2}
}
\approx 
\frac{c^{1/2}}{4 \sqrt{2} \sigma^{3/4} 
\sqrt{\omega / c - k} \, (k - \sigma^{1/4} \omega^{1/2})}
\approx
\label{eq9009}
\end{equation}
\[
\frac{c^{1/2}}{4 \sqrt{2} \i \sigma^{3/4} 
\sqrt{k- \omega / c} \, (k - \omega / (2 c) - c\sqrt{\sigma}/2)}.
\]

To estimate the integral, one can use the method described 
in Appendix~A1. This is the integral of the type Eq.~(\ref{A001})
with $\mu_1 = 1$, $\mu_2 = 1/2$, $\omega_\dag = \omega_*$, 
$k_\dag = k_*$, 
$v_1 = 2c$, $v_2 = c$. Formula Eq.~(\ref{A013}) can be used. The result is
\begin{equation}
u_3 (t,x) = \frac{
 \rho f_0 c^{3/2} V^{1/2} \exp \{ \i k_* x - \i \omega_* t + 3\pi \i/4 \}
}{
2\sqrt{2} \pi^{1/2} T \sigma^{3/4} x^{1/2} \sqrt{2c -V}
}.
\label{eq9010}
\end{equation}

\vskip 6pt
\noindent
{\bf Estimation of $u_{2c}$}

Assume that  $V \approx 2c$. Use the first approximation Eq.~(\ref{eq9009}). 
Apply the procedure of estimation described in 
Appendix~A3. The standard integral is Eq.~(\ref{A201})
with 
$\omega_\dag = \omega_*$, 
$k_\dag = k_*$, 
$v_1 = 2c$,
$v_2 = c$,
$\alpha = (8 \sigma^{1/2} c^3)^{-1}$. 

The estimation of the integral is given by Eq.~(\ref{A215}). The result is 
\begin{equation}
u_{2c} (t, x) =  
\frac{\i\rho f_0 c^{7/4}}{4(2)^{1/4} \pi T \sigma^{5/8} x^{1/4} }
\exp \{ \i k_* x - \i \omega_* t\} B \left(
\frac{
x^{1/2} \sigma^{1/4} (2c-V)
}{
 \sqrt{2} c^{1/2}
}
\right)
\label{eq9011}
\end{equation} 
with $B$ given by Eq.~(\ref{A213a}) and Eq.~(\ref{A212}).
 
The asymptotics Eq.~(\ref{eq3013}) and Eq.~(\ref{eq3014}) provide Eq.~(\ref{eq9004})
and Eq.~(\ref{eq9005}).

%%%%%%%%%%%%%%%%%%%%%%%%%%%%%%%%%%%%%%%%%%%
\subsection{Non-local estimation of Eq.~(\ref{eq1014}) for $V \approx c$}

Consider the case $V \approx c$. Our aim is to build an estimation 
of the integral over a long spot marked by ``0'' sign and located along the 
branch $k = \omega / c$ in Fig.~\ref{fig18a}, right. In other words, the aim is to compute 
$u_c$ from Eq.~(\ref{eq9003a}).

\begin{figure}[ht]
\centerline{\epsfig{file=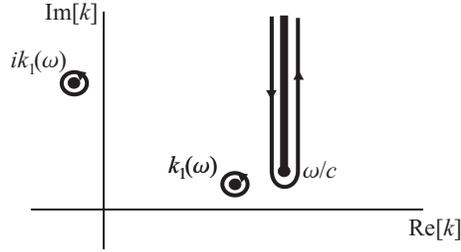, width = 6cm}}
\caption{Deformation of integration contour in the $k$-domain}
\label{fig13}
\end{figure}

Let be 
\begin{equation}
t = x/c + \delta t.
\label{eq1108}
\end{equation}
Change the order of integration in Eq.~(\ref{eq1014}) and take the integral with respect 
to $k$ for a fixed~$\omega$. For this, shift the integration contour into the upper half-plane
since the factor $e^{i k x}$ decays there. Note that
there are two poles and a branch point in the 
upper half-plane of~$k$. The poles are $k = k_1 (\omega) = \sigma^{1/4} \omega^{1/2}$
and 
$k = \i k_1 (\omega)$, while the branch point is $k = \omega / c$. 
After the deformation, the contour will look as shown in Fig.~\ref{fig13}.

%%%%%%%%%%%%%%%%%%%%%%%%%%%%
%%%%%%%%%%%%%%%%%%%%%%%%%%%%%%%%%%%%%%% 

The integral can be written as 
\begin{equation}
p(t , x, 0) = u_{\rm r1} (t, x) + u_{\rm r2} (t, x)  + 
u_{\rm b} (t, x),
\label{eq1101}
\end{equation}
where the first two terms are the residue integrals:
\begin{equation}
u_{\rm r1} (t, x) = -\frac{\rho f_0 c}{8 \pi T \sigma^{3/4}}
\int \limits_{- \infty + \i \epsilon}^{\infty + \i \epsilon}
\frac{\exp \{\i k_1 (\omega) x - \i \omega x/c - \i \omega \, \delta t )\} 
}{\sqrt{\omega - \omega_*}}
\rmd \omega,
\label{eq1102}
\end{equation}
\begin{equation}
u_{\rm r2} (t, x) = -\frac{\i \rho f_0 c}{8 \pi T \sigma^{3/4}}
\int \limits_{- \infty + \i \epsilon}^{\infty + \i \epsilon}
\frac{\exp \{- k_1 (\omega) x - \i \omega x/ c - \i \omega \, \delta t)\} 
}{\sqrt{\omega + \omega_*}}
\rmd \omega.
\label{eq1103}
\end{equation}
The branch point integral can be estimated for large $x$: 
\begin{equation}
u_{\rm b} (x, t)\approx \frac{\rho f_0 e^{\i \pi /4} c^{9/2}}{2 \sqrt{2} \pi^{3/2} T x^{1/2}}
\int \limits_{- \infty + \i \epsilon}^{\infty + \i \epsilon}
\frac{\exp \{- \i \omega \, \delta t\} }{(\omega^2 - \omega_*^2)\sqrt{\omega}}
\rmd \omega. 
\label{eq1104}
\end{equation}

\begin{figure}[ht]
\centerline{\epsfig{file=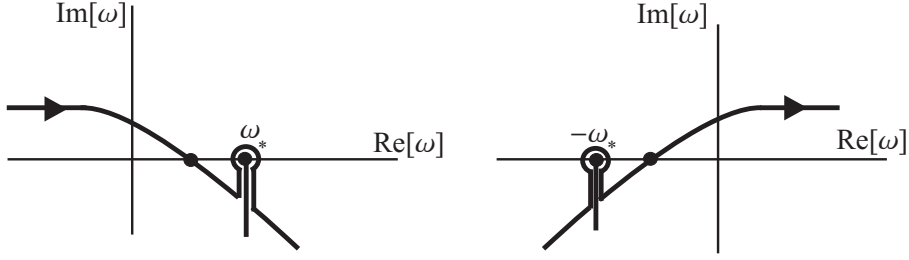, width = 12cm}}
\caption{Contour integration for estimation of 
$u_{\rm r1}$ (left) and $u_{\rm r2}$ (right) in the $\omega$-domain}
\label{fig12b}
\end{figure}

The contours for the first two integrals  can be deformed as it is shown in Fig.~\ref{fig12b} and estimated for large~$x$.
The result of the estimation for each integral is a sum of a saddle point term and of 
the branch cut term. We are interested only in the branch cut term, since the saddle 
point terms are equal to $u_1$ and $\bar u_1$ for $V = c$.
Denote the branch cut terms of $u_{\rm r1}$ and $u_{\rm r2}$ by  
$u_{\rm r1b}$ and $u_{\rm r2b}$, respectively. Their estimations are

%%%%%%%%%%%%%%%%%%%%%%%%%%%%
%%%%%%%%%%%%%%%%%%%%%%%%%%%%%%%%%%%%%%% 

\begin{equation}
u_{\rm r1b} (t, x) \approx
-\frac{\rho f_0 c^{3/2} \exp \{- \i \omega_* \delta t  - \i \pi /4\}}{
2 \sqrt{2} T \pi^{1/2} \sigma^{3/4} x^{1/2}
},
\qquad 
u_{\rm r2b} (t, x) = \bar u_{\rm r1b} (t, x) 
\label{eq1105}
\end{equation}

\begin{figure}[ht]
\centerline{\epsfig{file=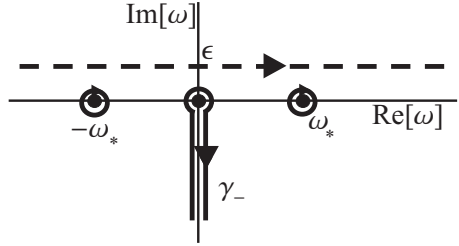, width = 6cm}}
\caption{Contour integration for estimation of 
Eq.~(\ref{eq1104}) in the $\omega$-domain}
\label{fig12c}
\end{figure}

Estimate the integral Eq.~(\ref{eq1104}). 
If $\delta t < 0$ the contour of integration can be closed in the upper half-plane, 
and $u_{\rm b} = 0$. 
If $\delta t > 0$ the contour should be shifted into the lower half-plane. 
The deformed contour is shown in Fig.~\ref{fig12c}. 
It consists of two polar terms and the branch term. 
Let the polar terms for the poles $\omega = \omega_*$ and $\omega = - \omega_*$
be $u_{\rm br1}$ and $u_{\rm br2}$, respectively. Let the branch cut integral
be denoted by $u_{\rm bb}$:
\begin{equation}
u_{\rm b} = u_{\rm br1} + u_{\rm br2} + u_{\rm bb}.
\label{eq1109}
\end{equation}  

%%%%%%%%%%%%%%%%%%%%%%%%%%%%
%%%%%%%%%%%%%%%%%%%%%%%%%%%%%%%%%%%%%%% 

A detailed computation yields 
\begin{equation}
u_{\rm br1}(t,x) = - u_{\rm r1b} (t,x),
\qquad  
u_{\rm br2}(t,x) = - u_{\rm r2b} (t,x),
\label{eq1110}
\end{equation}
\begin{equation}
u_{\rm bb} (t,x) = 
\frac{\rho f_0 e^{\i \pi /4} c^{9/2}}{2 \sqrt{2} \pi^{3/2} T x^{1/2}}
\int_{\gamma_-} 
\frac{\exp \{- \i \omega \, \delta t\} }{(\omega^2 - \omega_*^2)\sqrt{\omega}}
\rmd \omega. 
\label{eq1111}
\end{equation}
The last expression can be rewritten as 
\begin{equation}
u_{\rm bb} (t, x) = - 
\frac{\rho f_0 c^{3/2}}{ \sqrt{2} \pi^{3/2} T x^{1/2} 
\sigma^{3/4}}
E(\sigma^{1/2} c^2 \delta t),
\label{eq1112}
\end{equation}
where
\begin{equation}
E(a) = 
\int \limits_{0}^{\infty}
\frac{\exp \{- \xi a \} }{(\xi^2 + 1)\sqrt{\xi}}
d\xi. 
\label{eq1107}
\end{equation}   
Some properties of function $E(a)$ are displayed in Appendix~A4. 

Finally, 
\begin{equation}
u_{c} = \left\{ \begin{array}{ll}
u_{\rm r1b} + \bar u_{\rm r1b} , & t < x/c , \\
u_{\rm bb}                     , & t > x/c  
\end{array} \right.
\label{eq1113}
\end{equation}

From Eq.~(\ref{eq1105}) if follows that 
\[
u_{\rm r1b} (t,x)=  u_3(t,x) \qquad \mbox{for} \quad V = c.
\]
This means that there is no intermediate zone before the front $t = x / c$,
i.~e.\ the representation $p = u_1 + u_3 + \mbox{C.C.}$ remains valid 
for all $t < x/c$ even for small  $x/c - t$.  
Indeed, this guarantees Eq.~(\ref{eq9005c}).

The property Eq.~(\ref{A402}) can be used to establish validity of Eq.~(\ref{eq9005d}). 

The width of the intermediate zone is $\delta t \sim \sigma^{-1/2} c^{-2}$.
This agrees with Eq.~(\ref{eq9005b}). 

The property Eq.~(\ref{A403}) yields that the field $u_c$ is continuous at $t = x/c$. 
This feature clarifies the meaning of the asymptotics Eq.~(\ref{eq1112}). One can see that 
the asymptotics Eq.~(\ref{eq9008}) for $u_2$ is singular at $V = c$. However, the 
intermediate zone matches 
\redtext{
the term $u_2$ with the  pulse $u_3 + \bar u_3$} without 
making the field singular.

%%%%%%%%%%%%%%%%%%%%%%%%%%%%%%%%%%%%%%%%%%
\subsection{Taking into account absorption in the plate}

For the motivating air / ice / water problem it is natural to expect that the 
ice plate possesses some absorption. Here our aim to analyze the influence 
of this absorption on the asymptotic estimation of the integral. 
For simplicity, 
we use the model integral Eq.~(\ref{eq1014}) for this study.

The absorption of bending waves is modeled in the simplest way: 
the bending stiffness $T$ of the plate is assumed to have a small {\em negative\/} 
imaginary part: 
\begin{equation}
T  = T' - \i T''.
\label{eqG01}
\end{equation} 
Thus, the Young modulus of the plate has a small negative 
imaginary part, and all other physical parameters are real. 
Such a model is rather elementary, and it can describe correctly only the 
neighborhood of the coincidence point $(\omega_* , k_*)$. This model may be 
invalid for the other crossing points and for the low- and high-frequency components, 
but our aim is to demonstrate here the applicability of our method on the simplest possible example.  

The presence of the negative imaginary part of $T$ leads to a
small {\em positive\/} imaginary part of $\sigma$:
\begin{equation}
\sigma = \sigma' + \i \sigma''. 
\label{eqG02}
\end{equation}   
According to the dispersion diagram 
\begin{equation}
k^4 - \sigma \omega^2 = 0,
\label{eqG03}
\end{equation}
the value $k(\omega)$ for positive real $\omega$ has a positive imaginary part, 
and $\omega (k)$ for positive real $k$ has a negative imaginary part.
This corresponds to decaying waves of slightly different types.

The complex dispersion diagram $d_1^c$ still can be defined
for Eq.~(\ref{eqG03}), while the real 
dispersion diagram $d_1$ is empty. Thus, our 
asymptotic analysis should be slightly modified.  

The dispersion diagram $d_1^c$ possesses the following 
fundamental {\em energetic property\/} near $(\omega_*, k_*)$:
if $(\omega , k) \in d_1^c$ and $k$ is real, then $\omega$
has a negative imaginary part. Obviously, this property guarantees 
that energy decays in the system. 
To see that the property should be valid, one can consider a resonator 
made of the plate, having the length of $2 \pi / k$, and bearing the 
periodicity condition at the ends. The value $\omega$ is the eigenvalue of this 
resonator, and the negative imaginary part of $\omega$ corresponds 
to decay. 
As it follows from this property, $d_1^c$ does not intersect the initial 
integration manifold~$\Gamma_0$.  
 
Instead of the real dispersion diagram $d_1$, one can introduce the 
{\em approximately real dispersion diagram}, which is the set of points 
$(\omega , k)$ having small values of $|{\rm Im}[\omega]|$ and $|{\rm Im}[k]|$.  
 
Let us find the intersection point $(\omega_* ,k_*)$ as a crossing of $d_1^c$
and $d_2^c$, i.~e.\ a solution of the equations Eq.~(\ref{eqG03}) and 
$\omega^2 / c^2 - k^2 = 0$. 
Indeed, the solution is Eq.~(\ref{eq1012d}). However, now $\omega_*$ and $k_*$ both have 
positive imaginary parts. 

One can introduce the saddle points on $d_1^c$. Namely, since $d_1^c$ is an analytic set, 
the (complex) group velocity $v_{\rm gr}$ 
is defined by Eq.~(\ref{B001a}) at almost all points of~$d_1^c$. 
The saddle points are the points at which the group velocity is equal to the 
formal velocity $V$, i.~e.\ where Eq.~(\ref{B001aa}) is valid. 
Note that the group velocity at a saddle point is real, but the point itself can be 
complex.  

%We assume that the group velocity has a small imaginary part at the approximately real 
%dispersion diagram.  

The main statement of the asymptotic estimation procedure formulated in 
Subsection~\ref{sec_main} 
should be reformulated as follows: 
{\em 
For almost all values of $V$, the leading terms of the asymptotics of the 
double Fourier integral are produced by the fragments of the integration surface 
that are located in neighborhoods of either the 
saddle points on the  approximately real branches or the crossing points 
of the approximately real branches. 
} 

Note that now the leading terms of the asymptotics can be exponentially decaying
as $x \to \infty$ due to the energy absorption.  

The sketch of the proof of this statement is direct  modification of that from 
Subsection~\ref{sec_main}. The only new part that should be added is the deformation 
of $\Gamma_0$ in the case of 
\[
{\rm Re}[v_{\rm gr}]  = 0, 
\qquad 
{\rm Im}[v_{\rm gr}]  \ne 0. 
\]  
In this case, in the linear approximation, the intersection of $d_1^c$
with the plane ${\rm Re} [\omega] = \omega'$, ${\rm Re} [k] = k'$
is a point rather than a line. Thus, any contour not passing through this point will be an eligible deformation.    
 
The procedure of estimation of the integrals near the crossing points and the saddle points remains the same. Indeed, the formulae for terms $u_1$, $u_3$, and $u_{2c}$ also 
remain the same, 
but one should take into account that $\sigma$ is complex. 

The term $u_3$ represented as Eq.~(\ref{eq9010}) has an important feature for complex~$\sigma$. 
Since $\omega_*$ has a positive imaginary part, this term is exponentially 
{\em growing\/} as $t$ grows. This does not contradict to our consideration,
since the field should be decaying (and is decaying) as $x \to \infty$ with 
constant~$V$.
The exponential growth of $u_3$ can be explained physically. For some fixed $x$, 
the smaller is $t$, the bigger distance the signal traveled in the plate 
before being radiated into the gas. Since the plate has losses, the smaller $t$
correspond to the smaller signal.

%%%%%%%%%%%%%%%%%%%%%%%%%%%%%%%%%%%%%%%%%%%%%%%%%%%%%%%%

\section{Comments to Problem~3}

%\redtext{The whole section is added in the revised version}

\subsection{The preliminary analysis}

In this section we are analyzing Problem~3 and its solution Eq.~(\ref{eq1014a}). 
First, let us illustrate the influence of water on the bending waves in ice.
Take the thickness of ice equal to $h = 3$\,cm, and the physical parameters as
\[
E = 9\,\mbox{GPa}, \qquad 
\nu = 0.3, \qquad 
\rho_p = 900\,\mbox{kg/m}^3, \qquad 
c_w = 1500 \, \mbox{m/s} , \qquad
\]
\[
\rho_w = 1000 \, \mbox{kg/m}^3, \qquad 
c = 343\, \mbox{m/s} , \qquad
\rho = 1.3 \, \mbox{kg/m}^3.
\]
We assume that ice is lossless.  
 
\begin{figure}[ht]
\centerline{\epsfig{file=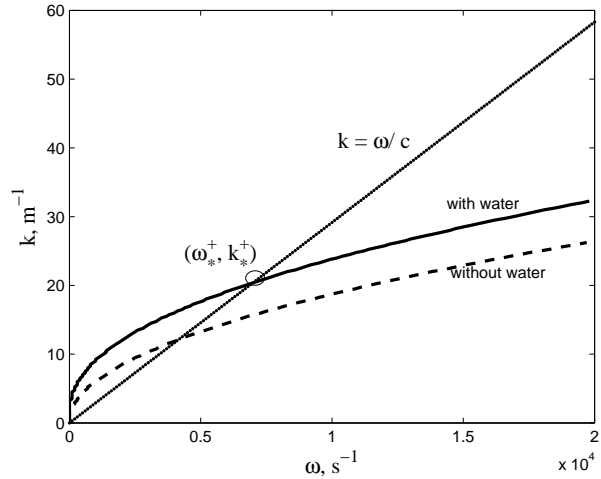, width = 9cm}}
\caption{Dispersion diagram for ice with $h = 3$\,cm loaded by water }
\label{dispersion_1}
\end{figure}

Fig.~\ref{dispersion_1} shows the dispersion diagrams for $D(\omega, k) = 0$
with $D$ defined as Eq.~(\ref{eqF01}). It is labeled as the curve ``with water'' in the graph. 
For a comparison, we plot also the set $d_1$ (the curve labeled as ``without water''),
and $d_2$ labeled as $k = \omega / c$. The coincidence point Eq.~(\ref{eqF02})
is indicated in the figure. One can see that the water affects the dispersion 
diagram of bending waves considerably, however qualitatively the diagrams are similar.

\begin{figure}[ht]
\centerline{\epsfig{file=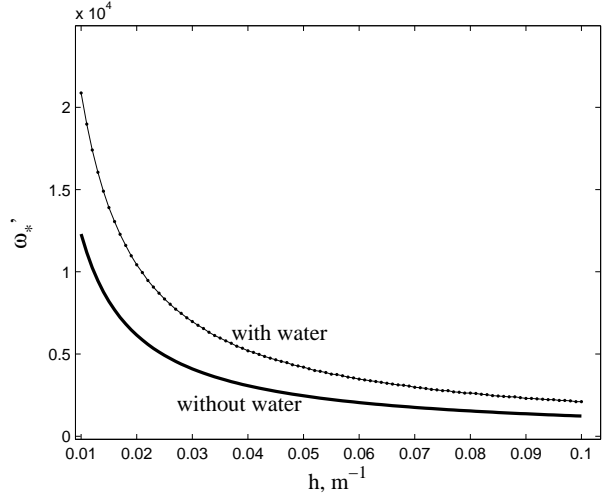, width = 9cm}}
\caption{The value of $\omega_*'$ as a function of $h$}
\label{dispersion_2}
\end{figure}

The coincidence frequencies $\omega_*\dag$ and $\omega_*$ of bending waves
have been computed 
for the range $h = 1 \dots 10$\,cm. The result is presented in Fig.~\ref{dispersion_2}.  
 
%%%%%%%%%%%%%%%%%%%%%%%%%%%%
%%%%%%%%%%%%%%%%%%%%%%%%%%%%%%%%%%%%%%% 

%%%%%%%%%%%%%%%%%%%%%%%%%%%%
%%%%%%%%%%%%%%%%%%%%%%%%%%%%%%%%%%%%%%% 

The system Eq.~(\ref{eqF02}) can be reduced to an algebraic equation. 
Let it be solved (say, numerically) for a given $h$, and let the values 
$\omega_*^\dag$ and $k_*^\dag = \omega_*^\dag / c$ be known. 
Then, one compute the derivatives 
\begin{equation}
\ptl_\omega D(\omega_*^\dag , k_*^\dag) = - 2 \sigma \omega_*^\dag -
\frac{\rho_w }{T} \frac{
2c^{-2} - 2c_w^{-2} + 1
}{(c^{-2} - c_w^{-2})^{3/2}} ,
\label{eqF09}
\end{equation} 
\begin{equation}
\ptl_k D(\omega_*^\dag , k_*^\dag) = 4 \frac{(\omega_*^\dag)^3}{c^3} + 
\frac{
\rho_w
}{T c (c^{-2} - c_w^{-2})^{3/2}} .
\label{eqF08}
\end{equation} 
The group velocity of the bending waves 
can be found according to the theorem of the implicit function: 
\begin{equation}
v_{\rm gr} = - \frac{\ptl_k D}{\ptl_\omega D}.
\label{eqF07}
\end{equation} 
Denote this value by $v_1$. 

A direct computation shows that the group velocity of the bending waves 
in the ice loaded by 
water is about $790\,\mbox{m/s}$
for the range 1\dots 10~cm, i.~e.\ is slightly higher than $2c$.
A good estimation is $v_1 \approx 2.3 c$ for the experiment.

%%%%%%%%%%%%%%%%%%%%%%%%%%%%%%%%%%%%%%%%%%%%%%%%%%%%%%%%%%%%%%%%%%%%%%%%%%%%

\subsection{Estimation of Eq.~(\ref{eq1014a}) near the intersection point}

The integral Eq.~(\ref{eq1014a}) doesn't have a form of a 2D Fourier integral, so it should be 
transformed to make our method applicable to it. Still, 
we assume that the analysis based on the locality remains valid. We are focusing here 
only on the terms related to the coincidence points $(\pm \omega_*^\dag , k_*^\dag)$. 

Assume that 
\begin{equation}
k_*^\dag L \gg 1.
\label{eqF10}
\end{equation}
In this case the Bessel function can be represented as 
\begin{equation}
J_0 (k L)  \approx \sqrt{\frac{1}{2 \pi k R}}
\left( 
\exp \{ \i k L - \i \pi /4\} 
+ 
\exp \{-\i k L + \i \pi /4\} 
\right).
\label{eqF11}
\end{equation}

Thus, the integral Eq.~(\ref{eq1014a}) near the points $(\pm \omega_*^\dag , k_*^\dag)$ can 
be written as 
\[
p(t, L , 0, 0) \approx  
\]
\begin{equation}
\frac{\i \rho f_0 (\omega_*^\dag)^3}{(2\pi)^{5/2} c T R^{1/2}}
\iint
\frac{
\exp \{ \i k L - \i \omega t - \i \pi /4\} 
+ 
\exp \{-\i k L - \i \omega t + \i \pi /4\} 
}{
\gamma (\omega , k) \, D(\omega , k)
}
\rmd k \, \rmd \omega. 
\label{eqF12}
\end{equation}

\begin{figure}[ht]
\centerline{\epsfig{file=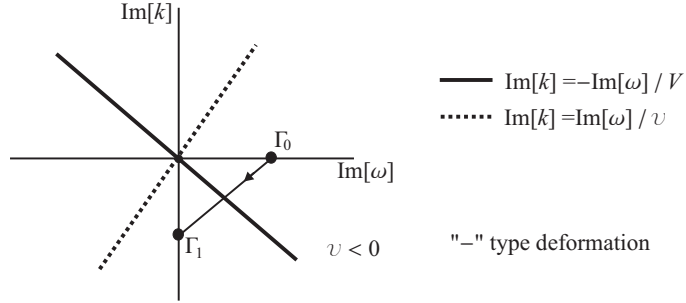, width = 9cm}}
\caption{``$-$'' type deformation for the second term of Eq.~(\ref{eqF12})}
\label{fig_add_def}
\end{figure}

Introduce the formal velocity by $V = L / t$.  
Let be $c < V < v_1$. 
Consider a neighborhood of the point 
$(\omega_*^\dag , k_*^\dag)$.
Formula Eq.~(\ref{eqF12}) shows that there are two terms in this 
integral. 
The first term corresponds to the deformation diagram shown in 
Fig.~\ref{fig18},~right. Thus, this term provides a non-decaying 
asymptotics. For the second term of Eq.~(\ref{eqF12}), one can select the 
``$-$'' type deformation shown in Fig.~\ref{fig_add_def} for 
both crossing branches. This means that the second term does not lead to a 
non-decaying asymptotics.
 
%%%%%%%%%%%%%%%%%%%%%%%%%%%%
%%%%%%%%%%%%%%%%%%%%%%%%%%%%%%%%%%%%%%% 

Thus, the task is reduced to estimation of the integrals 
\begin{equation}
p(t, L , 0, 0) \approx  
\frac{i \rho f_0 (\omega_*^\dag)^3}{(2\pi)^{5/2} c T L^{1/2}}
\iint_{\Omega_+}
\frac{
\exp \{ \i k L - \i \omega t - \i \pi /4\} 
}{
\gamma (\omega , k) \, D(\omega , k)
}
\rmd k \, \rmd \omega
+ 
\label{eqF13}
\end{equation}
\[
\frac{\i \rho f_0 (\omega_*^\dag)^3}{(2\pi)^{5/2} c T L^{1/2}}
\iint_{\Omega_-}
\frac{
\exp \{-\i k L - \i \omega t + \i \pi /4\} 
}{
\gamma (\omega , k) \, D(\omega , k)
}
\rmd k \, \rmd \omega, 
\]
where $\Omega_\pm$ are the fragments of the integration surface in the neighborhoods of 
$(\pm \omega_*^\dag , k_*^\dag)$. 

The resulting integrals have the form similar to those have been studied. They 
are double Fourier integrals whose denominators contain a crossing of a
quadratic branch set and a polar set.   
Thus, these integrals can be estimated by using the technique 
described above, providing the terms similar to $u_1$, $u_3$, and $u_{2c}$ of
Eq.~(\ref{eq9001}) and Eq.~(\ref{eq9003}). 

The non-local term similar to $u_{\rm bb}$ requires some special consideration , which is beyond the scope of the current paper. Still, we believe that its structure is qualitatively similar to what has been found above.

%%%%%%%%%%%%%%%%%%%%%%%%%%%%%%%%%%%%%%%%%%%%%%%%%%%%%%%%%%%%%%%%%%%%%%%%%%%

\subsection{Experimental results and their interpretation}

Let us demonstrate the experimental results obtained by one of the authors. 
The acoustic signals have been recorded for the parameters 
$h \approx 3$\,cm and $L = 155$\,m. Note that the thickness of ice is known very approximately, 
and we should admit that it could be varying over the lake surface due to natural reasons. 
  
\begin{figure}[ht]
\centerline{\epsfig{file=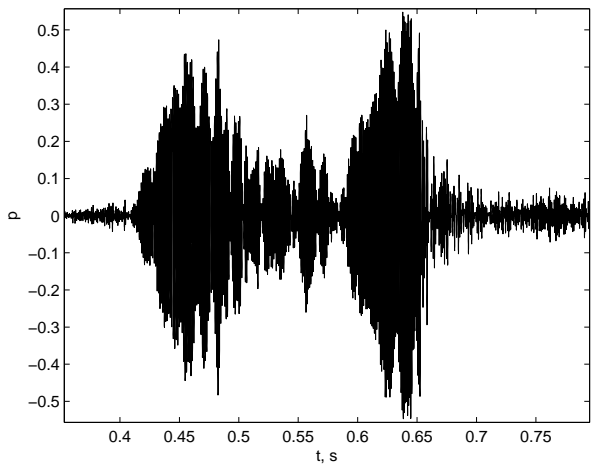, width = 9cm}}
\caption{Experimental signal for $h \approx 3$\,cm, $L = 155$\,m}
\label{signal}
\end{figure}

The shape of a typical signal is shown in Fig.~\ref{signal}. The zero mark of the time was 
not set properly, so the exact starting time of the signal is not known. 
We assume that, mainly, the signal caused by the coincidence point is visible 
(this is $u_3 + \bar u_3$ in Eq.~(\ref{eq9001})). 
The duration of the pulse is about 0.25\,s,
and this agrees very well with the rough estimation~$L/(2c)$.  
  
%%%%%%%%%%%%%%%%%%%%%%%%%%%%
%%%%%%%%%%%%%%%%%%%%%%%%%%%%%%%%%%%%%%% 

\begin{figure}[ht]
\centerline{\epsfig{file=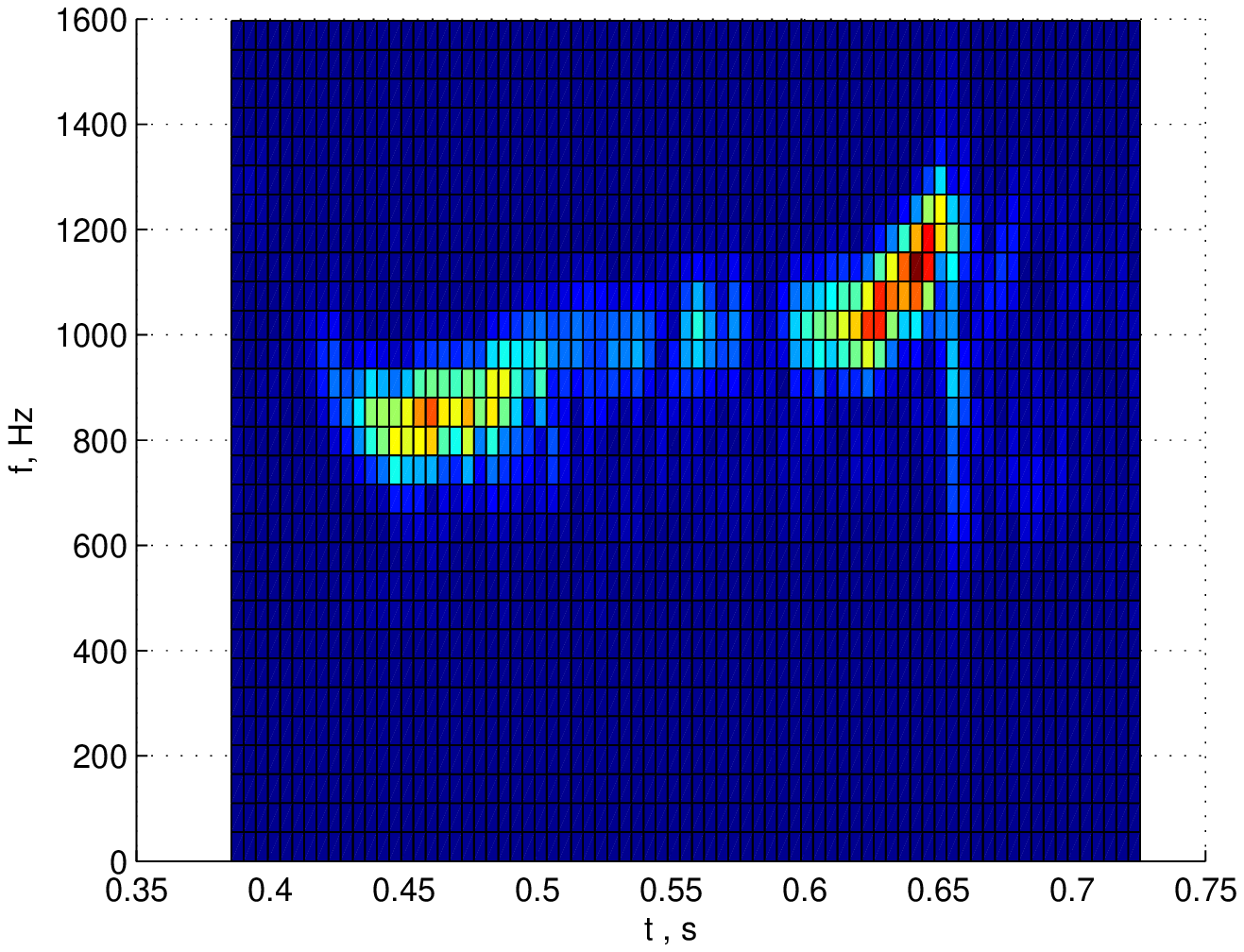, width = 8cm}
\epsfig{file=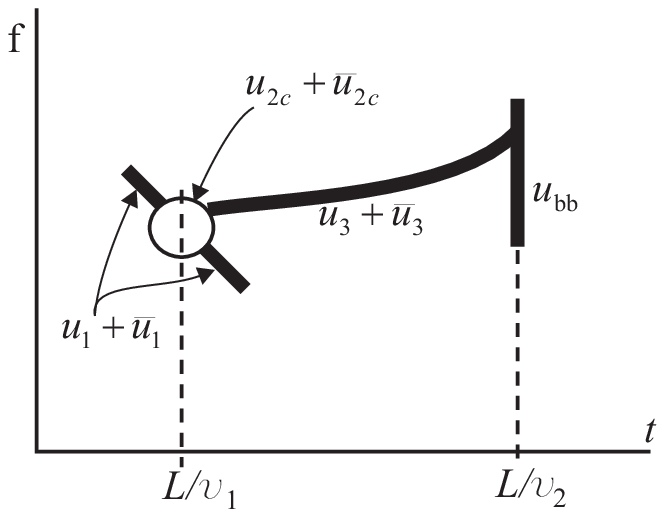, width = 7cm}
}
\caption{The spectrogram of the experimental signal and its interpretation}
\label{spectrogram}
\end{figure}

The spectrogram of the signal is shown in Fig.~\ref{spectrogram},~left. The spectrogram is obtained with short-time Fourier transform with a reasonable Gaussian window function. 
The vertical axis shows the frequency ${\rm f} = \omega / (2 \pi)$.   

%%%%%%%%%%%%%%%%%%%%%%%%%%%%
%%%%%%%%%%%%%%%%%%%%%%%%%%%%%%%%%%%%%%% 

Our interpretation of the spectrogram is shown in Fig.~\ref{spectrogram},~right.
The main component is $u_3 + \bar u_3$. It is almost monochromatic.
The frequency agrees with the thickness of ice.  
The reason of frequency elevation at the end of the signal is unclear, we can assume that 
the thickness of ice decreased closer to the shore.
The variations of the amplitude of this pulse are probably described by 
a sort of speckle pattern. 

At the start of the pulse one can see the fragment of the signal $u_1 + \bar u_1$.
This signal is strongly dispersive: the parts with higher frequencies go faster. 
That's why the frequency decays with time. An  estimation 
of the slope of the spectrogram is 
\begin{equation}
\frac{\rmd {\rm f}}{\rmd t} = - \frac{v_1^3 \sqrt{\sigma}}{4 \pi L}.
\label{eqF14}
\end{equation}
This estimation works reasonably well for the observed signal.

\begin{figure}[ht]
\centerline{\epsfig{file=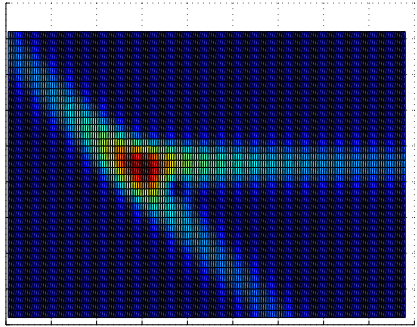, width = 7cm}
}
\caption{A model spectrogram of $u_{2c}$ (see Eq.~(\ref{eq9011}))}
\label{spectrogram_B}
\end{figure}

The zone of mixing of $u_1 + \bar u_1$ and $u_3 + \bar u_3$ should be described 
by the term $u_{2c} + \bar u_{2c}$ of Eq.~(\ref{eq9001}).
For illustration, we plot a model 
spectrogram of the function Eq.~(\ref{eq9011})
in Fig.~\ref{spectrogram_B}. The structure of the spectrogram is easily explained by the 
asymptotics Eq.~(\ref{eq3013}), Eq.~(\ref{eq3014}).   The inclined line 
corresponds to the first term of Eq.~(\ref{eq3013}) and to the term Eq.~(\ref{eq3014}). 
The straight line corresponds to the second term of Eq.~(\ref{eq3013}).

%%%%%%%%%%%%%%%%%%%%%%%%%%%%
%%%%%%%%%%%%%%%%%%%%%%%%%%%%%%%%%%%%%%% 

The back front of the pulse is visible as a vertical line in the spectrogram. 
This corresponds to a fast process in time. Our estimation of the
width of the 
pulse $u_{\rm bb}$ is $\sigma^{-1/2} c^{-2}$. The frequency band of such a pulse is 
about $1$~kHz, and this agrees with the spectrogram.    
    
The main problem for the interpretation is the dependence of the amplitude 
of the term $u_3 + \bar u_3$ 
vs.\ time. The asymptotic formula Eq.~(\ref{A013}), which remains valid 
for the term due to the coincidence point, yields the amplitude dependence 
$
~(v_1 t - L )^{-1/2},
$ 
which a decaying function for $t > L/v_1$. However, Fig.~\ref{signal}
shows that the back part of the pulse has the amplitude bigger than the front part. 
One possible explanation is the presence of decay in the ice. As we claimed above, the 
factor 
$\exp \{ - \i \omega_*^\dag t \}$ is exponentially {\em growing\/} in time if 
the ice is lossy. 
Another reason for the complicated behavior of amplitude of $u_3$ vs time
may be a refraction of sound wave near the surface of ice.  
 
To provide a detailed description of the experiment, one should
take into account 
that the altitude of the microphone was about $1.5$\,m height, not zero. 
The signal $u_1$ (the sound wave accompanying the bending wave in the ice)
strongly depends on the altitude of the microphone. The part with 
$x / t  < v_1$ corresponds to supersonic radiation of sound wave, so this component should be clearly visible in the recording. The part of $u_1$ with  
$x / t > v_1$ corresponds to the subsonic radiation, and this wave 
should be exponentially decaying in the air. The characteristic length is about the wavelegth, 
i.~e.\ about 30\,cm. 
However, although the subject seems to be interesting and practically important, 
the asymptotic estimation of the Fourier integral 
with $z \ne 0$ falls beyond the scope of the current paper.   
\color{black}

%%%%%%%%%%%%%%%%%%%%%%%%%%%%%%%%%%%%%%%%%%%%%%%%%%%%%%%%%%%%%%

\section{\redtext{Conclusions}}
 
The paper can be summarized as follows. The integrals Eq.~(\ref{eq1012}) and Eq.~(\ref{eq1014})
are analyzed. Both integrals describe non-stationary wave processes in two-component systems
with a high density contrast between the subsystems. The expressions Eq.~(\ref{eq1012}) and Eq.~(\ref{eq1014})
are 2D Fourier integrals. The denominators of the integrands have zero sets that are crossing. 

The paper contains an exact computation of Eq.~(\ref{eq1012}) and an asymptotic estimation 
of Eq.~(\ref{eq1014}). In both cases we find a monochromatic pulse existing for 
$x / (2c)<  t < x/ c$. The frequency and the wavenumber of this pulse are equal to those 
of the point of phase synchronism of the subsystems (the coincidence point).

A general scheme of estimation of a 2D Fourier integral is presented. The procedure of estimation is based on the fact (the ``main statement'') that the field components are produced by 
saddle points on the branches of the dispersion diagrams and by crossing points 
of the branches of the dispersion diagrams. A set of standard integrals
related to such points are given in Appendix. 

The ``main statement'' is formulated for a very particular case: the branches of the dispersion 
diagram are graphs of slowly varying functions.  Here ``slowly varying'' means that the size of a typical zone of variation 
of $d_1$ and $d_2$ is much bigger than the size of the domain of influence (DOI
in terms of \cite{Borovikov1994}) for a saddle point integral. Note that in a general case, the 
dispersion diagram contains avoiding crossing where the function changes in a small 
zone. Our ``main statement'' does not work in the general case, and it should be replaced
by a more sophisticated theorem.   
 
\redtext{
Although the methods developed in this paper are applicable to double Fourier integrals, 
we demonstrate how they can be applied to a more realistic 3D problem of sound generation by 
the ice layer loaded by the water substrate. Such a problem describes the experimental 
setting that is a motivation of the paper. 
The solution of the problem is described by a Fourier intergal in time and a
Fourier--Bessel integral in space. We demonstrate that some methods
of estimation
still can be applied to this integral.
}

\redtext{
The spectrogram of the signal is interpreted in terms of the asymptotic analysis. 
The main features of the experimental signal are 
explained correctly by our analysis, namely the frequency of the 
main signal, its duration, the structure of the front and the back of the pulse.       
}

The work can be continued in four directions. First, one can consider 
avoiding--crossings instead of crossings of the dispersion diagrams. As it is known,
 this is a more realistic situation emerging when the contrast between 
subsystems is not very high. Second, one can study the field at the observation point 
not close to the horizontal surface. This leads to appearance of the factor 
$\exp \{ i \gamma z \}$, which makes the estimation procedure different. Third, 
one can introduce and study more of the standard integrals. For example, finding the 
asymptotics of Eq.~(\ref{eq1014}) for $V/ c \approx 0$ requires a standard integral with 
two square root singularities, two polar singularity and a double zero.  
Fourth, the deformations of the integration manifold described here can be used to develop 
efficient methods of numerical computation of double Fourier integrals.   
   
\section*{\redtext{Acknowledgements}}   
   
AVS thanks Prof.~C.J.~Chapman for fruitful discussions of the waveguide subjects during 
the INI programme WHT: 
``Bringing pure and applied analysis together via the Wiener--Hopf technique, its generalizations and applications''.
The WHT programme was supported by EPSRC (grant no. EP/R014604/1). The visit of AVS
to the INI program and his work on non-local asymptotical expansion has been partly supported by the grant from the Simon's foundation.   
%\redtext{
%The authors are grateful to Prof. A.V.~Metrikine for a kind criticism that enabled us 
%to make the paper better.  
%}   
   
The study of standard integrals has been funded by RFBR, project number 19-29-06048.  
 
\newpage
\bibliography{IceBib}
\bibliographystyle{unsrt}

\begin{appendices}
\titleformat{\section}{\normalfont\Large\bfseries}{\appendixname~\thesection.}{1em}{}
\numberwithin{equation}{section}
\renewcommand\thefigure{\thesection.\arabic{figure}}
\setcounter{figure}{0}  
%%%%%%%%%%%%%%%%%%%%%%%%%%%%%%%%%%%%%
\section{Standard local integrals}
\label{AppendixA}
\subsection{Crossing of two singular sets}

Consider the integral 

\begin{equation}
I (x, V) = 
\int \limits_{-\infty}^{\infty}
\int \limits_{-\infty + \i \epsilon}^{\infty + \i \epsilon}
\frac{
\exp \{ \i x (k - \omega / V) \} 
}{
(k - \psi_1 (\omega))^{\mu_1} (k - \psi_2 (\omega))^{\mu_2} 
}  \rmd \omega \, \rmd k 
\label{A001}
\end{equation}
with  
\begin{equation}
\psi_{1,2} (\omega) = k_\dag + \frac{\omega - \omega_\dag}{v_{1,2}},
\label{A002}
\end{equation}   
where $\omega_\dag$, $k_\dag$, $v_1$, $v_2$ are some real values.
%As usually, $V = x / t$. 
One can see that the integrand of Eq.~(\ref{A001}) has two singular sets (lines): 
\[
k = \psi_{1,2}(\omega) .
\]
The lines are crossing at the point $(\omega_\dag, k_\dag)$.
The values $v_1$ and $v_2$ are group velocities of 
the ``dispersion diagrams'' $k = \psi_{1,2} (\omega)$ at the crossing point. 
Let be $v_1 > v_2$. We do not assume that $v_1$ and $v_2$ are positive. 

Let us find the asymptotics of $I$ as $V = \mbox{const} >0$ and $x \to \infty$.

Real parameters $\mu_{1,2}$ determine the type of singularities. We are particularly interested in the cases $\mu = 1$ (a polar set), or $\mu = 1/2$  (a branching with integrable singularity). 
Assume for definiteness 
that the function $(\cdot)^{\mu}$ is 
positive real if the argument is positive real, and that this function is continuous on the integration surface.    

%Note that the integrand of Eq.~(\ref{A001}) is a holomorphic function of 
%complex variables $\omega$ and $k$
%everywhere except the singular sets. 
%Thus, according to the 2D Cauchy's theorem \cite{Shabat2} 
%one can deform the surface of integration surface without changing the
%value of the integral. 
%However, here we can avoid using the formalism of differential forms and 2D complex analysis 
%staying in the framework of repeated integrals and 1D contour integration.  

Introduce the variables 
\begin{equation}
\eta_{1,2} = k - \psi_{1,2} (\omega) = (k - k_\dag) - \frac{\omega - \omega_\dag}{v_{1,2}}.
\label{A003}
\end{equation}
The integral $I$ can be rewritten as 
\begin{equation}
I (x , V) = 
\frac{v_1 v_2 \exp\{ \i x (k_\dag - \omega_\dag / V) \}
}{v_1 - v_2}
\times \qquad \qquad \qquad \qquad \qquad \qquad \qquad \qquad 
\label{A004}
\end{equation}
\[
\int \limits_{-\infty - \i \epsilon / v_1}^{\infty - \i \epsilon / v_1}
\exp \left\{ 
 \frac{\i x ( V - v_2) v_1 \eta_1 }{V(v_1 - v_2)} 
\right\}
\frac{\rmd\eta_1 }{\eta_1^{\mu_1}} 
\int \limits_{-\infty - \i \epsilon / v_2}^{\infty - \i \epsilon / v_2}
\exp \left\{ 
 \frac{\i x (v_1 - V) v_2 \eta_2 }{V(v_1 - v_2)} 
\right\}
\frac{\rmd \eta_2}{ \eta_2^{\mu_2}} .
\]

\begin{figure}[ht]
\centerline{\epsfig{file=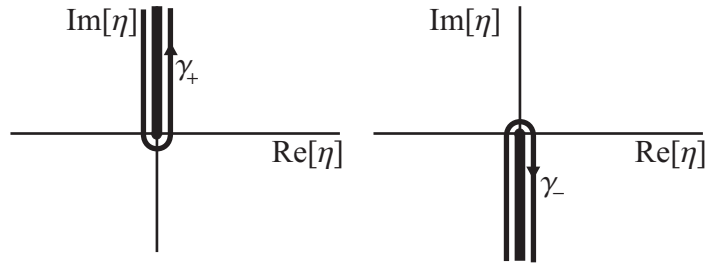}}
\caption{Deformed contours of integration in the $\eta$-plane}
\label{fig15}
\end{figure}

Note that the integration contour in the plane $\eta_j$, $j = 1,2$,
passes below the real axis if $v_j > 0$ and above the real axis if $v_j < 0$. 
Using this fact, close the contours of integration in appropriate half-planes 
and obtain that 
\begin{equation}
I (x, V) = 0 \quad \mbox{if} \quad V < v_2 \quad \mbox{or} \quad V > v_1.
\label{A005}
\end{equation}
If $v_2 < V < v_1$, then 
\[
I (x, V) = 
\]
\begin{equation}
\frac{v_1 v_2 \exp\{ \i x (k_\dag - \omega_\dag / V) \}}{v_1 - v_2}
I_{\rm c} \left( \mu_1,  \frac{x (V - v_2) v_1}{(v_1 - v_2)V} \right) 
I_{\rm c} \left( \mu_2,  \frac{x (v_1 - V) v_2}{(v_1 - v_2)V} \right),
\label{A007}
\end{equation}
\begin{equation}
I_{\rm c}(\mu,a) \equiv
\int_\gamma \frac{\exp\{ \i a \xi \}}{\xi^\mu} \rmd\xi,
\label{A007a}
\end{equation}
where $\gamma = \gamma_+$ if $a>0$ and $\gamma = \gamma_-$ if $a <0$.
Contours $\gamma_{+}$ and $\gamma_-$ are shown in Fig.~\ref{fig15}. 

%%%%%%%%%%%%%%%%%%%%%%%%%%%%
%%%%%%%%%%%%%%%%%%%%%%%%%%%%%%%%%%%%%%% 

Compute $I_{\rm c} (\mu, a)$. If $\mu$ is not a positive integer then 
\begin{equation}
I_{\rm c} (\mu, a) = e^{\pi \i (1-\mu)/2} (1 - e^{2\pi \i \mu })
a^{\mu - 1} \Gamma(1-\mu) \qquad \mbox{if} a >0,
\label{A008}
\end{equation}
\begin{equation}
I_{\rm c} (\mu, a) = e^{\pi \i (\mu-1)/2} (1 - e^{-2\pi \i \mu})
(-a)^{\mu - 1} \Gamma(1-\mu) \qquad \mbox{if } a <0,
\label{A009}
\end{equation}
$\Gamma(\cdot)$ is the Gamma-function.

If $\mu$ is a positive integer than 
\begin{equation}
I_{\rm c} (\mu, a) = \frac{2\pi e^{\pi \i \mu/2} a^{\mu - 1}}{(\mu -1)!}
\qquad \mbox{if } a >0,
\label{A010}
\end{equation}
\begin{equation}
I_{\rm c} (\mu, a) = - \frac{2\pi e^{\pi \i \mu/2} a^{\mu - 1}}{(\mu -1)!}
\qquad \mbox{if } a < 0.
\label{A011}
\end{equation}

In particular, for positive $a$
\begin{equation}
I_{\rm c} (1, a) = 2\pi \i,
\label{A107}
\end{equation}
\begin{equation}
I_{\rm c} (1/2, a) = \frac{2e^{\pi \i /4}\sqrt{\pi}}{\sqrt{a}} .
\label{A107a}
\end{equation}

Let be $v_1 > V >v_2 > 0$, $\mu_1 = \mu_2 = 1$. Then
\begin{equation}
I(x, V) = \frac{4 \pi^2 v_1 v_2 \exp \{ \i x (k_\dag - \omega_\dag / V)\} }{v_2 - v_1} .
\label{A012}
\end{equation}
Let be $v_1 > V >v_2 > 0$, $\mu_1 = 1$, $\mu_2 = 1/2$. Then
\begin{equation}
I(x, V) = - \frac{  
4 \pi^{3/2} v_1 v_2^{1/2} V^{1/2}\exp\{ \i x (k_\dag - \omega_\dag / V) - \pi \i /4 \}
}{
x^{1/2} (v_1 - v_2)^{1/2} 
   (v_1 - V)^{1/2}
}.
\label{A013}
\end{equation}
Finally, let be 
$v_2 < 0 < V < v_1$, $\mu_1 = \mu_2 = 1/2$. Then
\begin{equation}
I(x , V) = \frac{
4 \pi  \sqrt{- v_1 v_2} \,V \exp\{ \i x (k_\dag - \omega_\dag / V) \}
}{
x  \sqrt{(v_1 - V)(V - v_2)} 
}.
\label{A014}
\end{equation}

%%%%%%%%%%%%%%%%%%%%%%%%%%%%%%%%%%%%%%%

\subsection{A saddle point on a singular set}

Consider the integral 
\begin{equation}
I(x, V) = \int \limits_{-\infty}^{\infty} 
\int \limits_{-\infty + \i \epsilon}^{\infty + \i \epsilon}
\frac{\exp \{ \i x (k - \omega / V) \}}{(k - \psi (\omega))^\mu} \rmd \omega \, \rmd k ,
\label{A101}
\end{equation}
where 
\begin{equation}
\psi (\omega) = k_{\rm s} + \frac{\omega - \omega_{\rm s}}{V} -
\alpha (\omega - \omega_{\rm s})^2 .
\label{A102}
\end{equation}
Parameters $\mu$, $\omega_{\rm s}$, $k_{\rm s}$, $\alpha$ are real. Parameter $\epsilon$
is small. One can see that the point $(\omega_{\rm s} , k_{\rm s})$
is a saddle point of the integral on the dispersion diagram $k = \psi (\omega)$. 
Indeed, 
\[ 
\alpha = - 2 \frac{\rmd^2 \psi (\omega_{\rm s})}{\rmd^2 \omega_{\rm s}} .
\] 

Introduce the coordinates 
\begin{equation}
\eta = k - \psi (\omega) = (k - k_{\rm s}) - \frac{\omega - \omega_{\rm s}}{V}
+ \alpha (\omega - \omega_{\rm s})^2,
\qquad 
\xi = \omega - \omega_{\rm s}.
\label{A103}
\end{equation}
After a deformation of the integration surface, obtain
\begin{equation}
I(x, V) = 
\exp\{\i x (k_{\rm s} - \omega_{\rm s} / V) \}
\int \limits_{-\infty }^{\infty }
\int_{\gamma_+}
\frac{\exp \{ \i x (\eta - \alpha \xi^2 ) \}
}{\eta^\mu} \rmd\eta \, \rmd \xi  .
\label{A104}
\end{equation}
The integral can be taken:
\begin{equation}
I(x, V) = \exp\{\i x (k_{\rm s} - \omega_{\rm s} / V) \} 
I_{\rm c} (\mu , x) I_{\rm a} (\alpha x), 
\label{A105}
\end{equation}
where 
\begin{equation}
I_{\rm a} (\xi) = 
\left\{ \begin{array}{ll}
\exp \{ -\i \pi / 4 \} \sqrt{\pi / \xi} , & \xi > 0, \\
\exp \{ \i \pi / 4 \} \sqrt{-\pi / \xi} , & \xi < 0.
\end{array} \right.
\label{A106}
\end{equation}

%%%%%%%%%%%%%%%%%%%%%%%%%%%%%

\subsection{ A saddle point near a crossing point of singular sets}

Consider the integral 
\begin{equation}
I(x , V) = 
\int \limits_{-\infty}^{\infty} 
\int \limits_{-\infty + \i \epsilon}^{\infty + \i \epsilon}
\frac{\exp \{ \i x (k - \omega / V)\} }{
(k - \psi_1 (\omega))^{\mu_1} (k - \psi_2 (\omega))^{\mu_2}
} \rmd \omega \, \rmd k ,
\label{A201}
\end{equation}
\begin{equation}
\psi_1 (\omega) = k_\dag + \frac{\omega - \omega_{\dag}}{v_1} 
- \alpha (\omega - \omega_\dag)^2,
\label{A202}
\end{equation}
\begin{equation}
\psi_2 (\omega) = k_\dag + \frac{\omega - \omega_\dag}{v_2} ,
\label{A203}
\end{equation}
Let be $v_1 > v_2 >0$. 
Assume that $V \approx v_1$, i.~e.\ the saddle point belongs to the branch 
$k = \psi_1 (\omega)$. 

\begin{figure}[ht]
\centerline{\epsfig{file=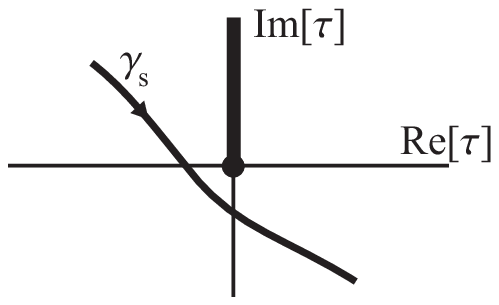}}
\caption{Contour $\gamma_{\rm s}$}
\label{fig09a}
\end{figure}

Introduce the variables 
\begin{equation}
\eta_1 = k - \psi_1(\omega) = (k - k_\dag) - \frac{\omega - \omega_\dag}{v_1}+
\alpha (\omega - \omega_\dag)^2,
\label{A204}
\end{equation}
\begin{equation}
\eta_2 = k - \psi_2(\omega) = (k - k_\dag) - \frac{\omega - \omega_\dag}{v_2}.
\label{A205}
\end{equation}
Solve the equations 
\[
\eta_1 = \delta k - \frac{\delta \omega}{v_1} + \alpha (\delta \omega)^2,
\qquad
\eta_2 = \delta k - \frac{\delta \omega}{v_2}
\]
with respect to the variables $\delta \omega = \omega - \omega_\dag$
and $\delta k = k - k_\dag$.
To get convenient formulae, assume that the term $\alpha (\delta \omega)^2$
is small, and the equation can be solved iteratively. After the second iteration
obtain 
\begin{equation}
\delta k \approx \frac{v_1 \eta_1 -v_2 \eta_2}{v_1 - v_2}
- \alpha \frac{v_1^3 v_2^2}{(v_1-v_2)^3} (\eta_1 - \eta_2)^2,
\label{A206}
\end{equation} 
\begin{equation}
\delta \omega \approx \frac{v_1v_2}{v_1 - v_2}(\eta_1 - \eta_2)
- \alpha \frac{v_1^3 v_2^3}{(v_1-v_2)^3} (\eta_1 - \eta_2)^2.
\label{A207}
\end{equation}
Using these approximations, 
write the exponential factor of $I$ in the form 
\begin{equation}
\exp \left\{ 
\i x \left( k - \frac{\omega}{V} \right)
\right\} \approx 
\exp \left\{ 
\i x \left( k_\dag - \frac{\omega_\dag}{V} \right)
\right\} \times \qquad \qquad \qquad \qquad 
\label{A208}
\end{equation}
\[
\exp 
\left\{    
\frac{
\i x (V - v_2) v_1
}{
(v_1 - v_2) V
} \eta_1
+
\frac{
\i x (v_1 - V) v_2
}{
(v_1 - v_2) V
} \eta_2
- 
\frac{\i \alpha x v_1^3 v_2^2 (V-v_2)}{(v_1 - v_2)^3V} (\eta_1-\eta_2)^2
\right\}.
\]
Since $V \approx v_1$, the factor $v_1 - V$ in the second term is small, and 
the size of the integration domain in $\eta_2$ is much bigger than the size in $\eta_1$. 
Thus, one can replace $(\eta_1-\eta_2)^2$ by $\eta_2^2$: 
\begin{equation}
\exp \left\{ 
\i x \left( k - \frac{\omega}{V} \right)
\right\} \approx 
\exp \left\{ 
\i x \left( k_\dag - \frac{\omega_\dag}{V} \right)
\right\} \times 
\label{A209}
\end{equation}
\[
\exp 
\left\{    
\frac{
\i x (V - v_2) v_1
}{
(v_1 - v_2) V
} \eta_1
+
\frac{
\i x (v_1 - V) v_2
}{
(v_1 - v_2) V
} \eta_2
-  
\frac{\i \alpha x v_1^3 v_2^2 (V-v_2)}{(v_1 - v_2)^3V} \eta_2^2
\right\}.
\]
The integral $I$ can be written approximately as
\[
I(x , V) \approx \frac{v_1 v_2 \exp \{\i x(k_\dag - \omega_\dag / V) \}}{v_1 - v_2}\times
\] 
\begin{equation}
I_{\rm c} \left(\mu_1, \frac{x (V-v_2) v_1}{(v_1 - v_2)V} \right)
I_{\rm s} \left(\mu_2, \frac{x (v_1 - V) v_2}{(v_1 - v_2)V} , 
\frac{\alpha x v_1^3 v_2^2 (V-v_2)}{(v_1-v_2)^3V} \right),
\label{A210}
\end{equation}
where 
\begin{equation}
I_{\rm s}(\mu, a , b) = \int_{\gamma_{\rm s}} 
\frac{\exp \{ \i a \xi - \i b \xi^2\}}{\xi^{\mu}} \rmd\xi
=
b^{(\mu - 1)/2} \hat I_{\rm s}(\mu , a /\sqrt{b})
,
\label{A211}
\end{equation}
\begin{equation}
\hat I_{\rm s}(\mu , \xi) =
\int_{\gamma_{\rm s}} \frac{\exp \{ \i (\tau \xi - \tau^2) \} }{\tau^\mu} \rmd\tau,
\label{A212}
\end{equation}
where the contour of integration of integration is shown in Fig.~\ref{fig09a}.

%%%%%%%%%%%%%%%%%%%%%%%%%%%%
%%%%%%%%%%%%%%%%%%%%%%%%%%%%%%%%%%%%%%% 

Function $\hat I_{\rm s}$ is rather complicated. We consider two cases.
For $\mu = 1$
\begin{equation}
\hat I_{\rm s} (1, \xi) = 2\pi \i(1- C(\xi/2 )),
\label{A213}
\end{equation} 
where $C$ is the Fresnel integral 
\begin{equation}
C(\xi) = \frac{1}{\sqrt{\pi \i}} 
\int \limits_{\xi}^{\infty} e^{\i \zeta^2} \rmd\zeta .
\label{eq2029}
\end{equation}
The well-known formula Eq.~(\ref{A213}) can be proven by differentiation of 
Eq.~(\ref{A212}) with respect to~$\xi$.

The Fresnel integral Eq.~(\ref{eq2029}) has asymptotics for real $a$ 
\begin{equation}
C(\xi) = \frac{\exp \{ \i \xi^2 + \i \pi /4 \}}{2\sqrt{\pi} \xi},
\quad \mbox{for} \quad 
\xi \gg 1,
\label{eq2029z}
\end{equation}
\begin{equation}
C(-\xi) = -\frac{\exp \{ \i \xi^2 + \i \pi /4 \}}{2\sqrt{\pi} \xi}  + 1,
\quad \mbox{for} \quad 
\xi \gg 1.
\label{eq2029y}
\end{equation}

\begin{figure}[ht]
\centerline{\epsfig{file=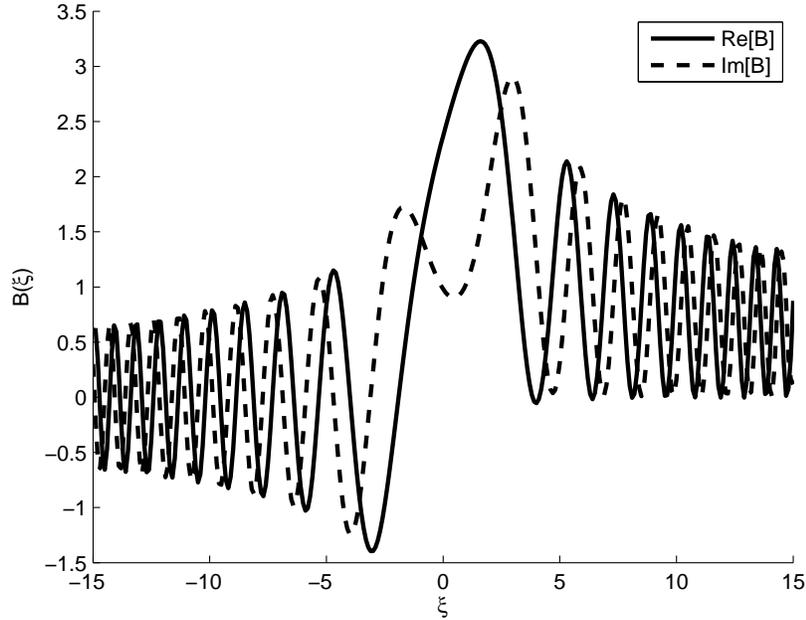, width=12cm}}
\caption{Numerically computed function $B(\xi)$}
\label{fig10}
\end{figure}

If $\mu =1/2$ the function 
\begin{equation}
B(\xi) \equiv \hat I_{\rm s} (1/2 , \xi) 
\label{A213a}
\end{equation} 
can be expressed through the functions of the parabolic cylinder \cite{Borovikov1994}. 
However, it is simple to tabulate $B(\xi)$ directly by using the definition Eq.~(\ref{A212}). The real and imaginary part of the function computed numerically are plotted in Fig.~\ref{fig10}.

%%%%%%%%%%%%%%%%%%%%%%%%%%%%
%%%%%%%%%%%%%%%%%%%%%%%%%%%%%%%%%%%%%%% 

Function $B(\xi)$ has the following asymptotics for real $\xi$: 
\begin{equation}
B(\xi) \approx \sqrt{\frac{2\pi}{\xi}} \exp \{ \i (\xi^2  - \pi ) /4\}
+ 
2 \sqrt{\frac{\pi}{\xi}} e^{\pi \i /4}
\quad \mbox{for} \quad 
\xi \gg 1,
\label{eq3013}
\end{equation}
\begin{equation}
B(-\xi) \approx \sqrt{\frac{2\pi}{\xi}} \exp \{ \i (\xi^2  +\pi ) /4\}
\quad \mbox{for} \quad 
\xi \gg 1,
\label{eq3014}
\end{equation}
The asymptotics Eq.~(\ref{eq3013}) and Eq.~(\ref{eq3014}) describe  function 
$B(\xi)$ well enough from $|\xi| \sim 3$. The maximum value of $B(\xi)$ is equal 
to~3.727.

Finally, an approximation of Eq.~(\ref{A201}) for $\mu_1 = \mu_2 = 1$ 
can be obtained by combining Eq.~(\ref{A210}), Eq.~(\ref{A213}), and Eq.~(\ref{A107})
and taking $V \approx v_1$:  
\begin{equation}
I(x, V) \approx \frac{4\pi^2 v_1 v_2
\exp \{ \i x (k_\dag - \omega_\dag / V) \}
}{
v_1-v_2
}
C\left(
\frac{
x^{1/2}  (V - v_1)
}{
2 \alpha^{1/2} v_1^{2} 
}
\right). 
\label{A214}
\end{equation}
Similarly, for $\mu_1 = 1$, $\mu_2 = 1/2$ obtain the approximation
\begin{equation}
I(x, V) \approx \frac{2\pi \i}{(\alpha x)^{1/4}}\sqrt{\frac{v_1v_2}{v_1 - v_2}}
\exp \{ \i x (k_\dag - \omega_\dag / V) \}
B\left(
\frac{
x^{1/2}  (v_1 - V)
}{
\alpha^{1/2} v_1^{2} 
}
\right).
\label{A215}
\end{equation}

\subsection{A special function for the non-local estimation}

\begin{figure}[ht]
\centerline{\epsfig{file=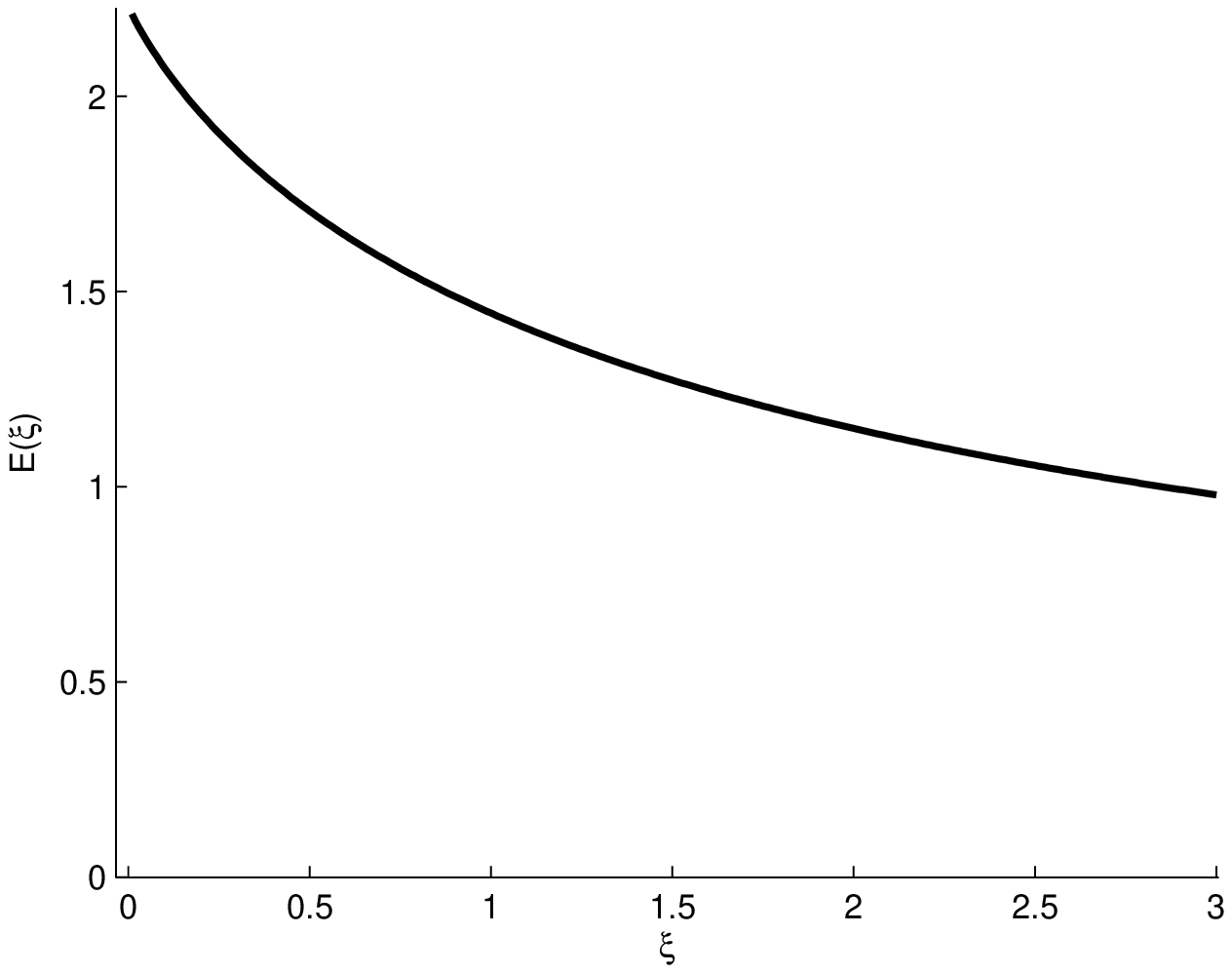,width = 12cm}}
\caption{Numerically computed function $E(\xi)$}
\label{fig22}
\end{figure}

Consider the integral defined for real $\xi > 0$
\begin{equation}
E (\xi) = \int \limits_0^{\infty}
\frac{e^{- \xi \tau}}{ (1+ \tau^2) \sqrt{\tau}} \rmd\tau.  
\label{A401}
\end{equation}
The graph of this function is shown in Fig.~\ref{fig22}. 

%%%%%%%%%%%%%%%%%%%%%%%%%%%%
%%%%%%%%%%%%%%%%%%%%%%%%%%%%%%%%%%%%%%% 

Function $E$ has the following asymptotics for $\xi \gg 1$:
\begin{equation}
E(\xi) \approx \frac{\sqrt{\pi}}{\sqrt{\xi}} .
\label{A402}
\end{equation}
We also note that 
\begin{equation}
E(0) = \frac{\pi}{\sqrt{2}}.
\label{A403}
\end{equation}    
    
\end{appendices}   

\end{document}